# Unboxing mutations: Connecting mutation types with evolutionary consequences

Emma L. Berdan[1*], Alexandre Blanckaert[2], Tanja Slotte[1], Alexander Suh[3,4], Anja M. Westram[5,6], and Inês Fragata[7*]

1. Department of Ecology, Environment and Plant Sciences, Science for Life Laboratory, Stockholm University, Stockholm, Sweden SE-10691
2. Laboratory of Genetics, University of Wisconsin-Madison, Madison, WI 53706
3. School of Biological Sciences – Organisms and the Environment, University of East Anglia, Norwich, United Kingdom, NR47TJ
4. Department of Organismal Biology – Systematic Biology, Uppsala University, Science for Life Laboratory, Uppsala, Sweden SE-75236
5. IST Austria, Am Campus 1, 3400 Klosterneuburg, Austria
6. Faculty of Biosciences and Aquaculture, Nord University, Bodø, Norway, N-8049
7. cE3c – Centre for Ecology, Evolution and Environmental Changes, Faculdade de Ciências, Universidade de Lisboa, Lisboa, Portugal

* - Corresponding author, emma.berdan@gmail.com, emmaberdan.weebly.com, irfragata@fc.ul.pt, @ines_fragata

## Abstract

Mutations are typically classified by their effects on the nucleotide sequence and by their size. Here, we argue that if our main aim is to understand the effect of mutations on evolutionary outcomes (such as adaptation or speciation), we need to instead consider their population genetic and genomic effects, from altering recombination rate to modifying chromatin. We start by reviewing known population genetic and genomic effects of different mutation types and connect these to the major evolutionary processes of drift and selection. We illustrate how mutation type can thus be linked with evolutionary outcomes and provide suggestions for further exploring and quantifying these relationships. This reframing lays a foundation for determining the evolutionary significance of different mutation types.

---

## Highlights

Occurrence rates vary greatly between different types of mutation (such as a transposable element vs. an inversion). There is an urgent need for more mutation rate studies in a wide variety of taxa to fully understand their relative contributions to evolutionary processes such as adaptation and speciation.

Mutations can affect multiple population genetic and genomic parameters such as recombination, effective population size, and methylation/chromatin state.

Merging molecular genetics with population and quantitative genetics allows us to connect mutation types with evolutionary outcomes.

---

## Introduction

Evolution depends on two major processes: (1) The random generation of genetic variation through mutation and its shuffling via recombination, and (2) subsequent changes in allele frequency through gene flow, genetic drift, and natural selection [1]. There are a wide variety of mutation types, which are often grouped by the change to the DNA sequence and size, ranging from point mutations to large structural variants [2]. However, from an evolutionary viewpoint, the most important characteristics of a mutation are its population genetic and genomic effects and how these may influence downstream evolutionary outcomes. Genetic studies have revealed that various types of mutations have drastically different population genetic and genomic effects ranging from changes in recombination rate to modifications of chromatin state. Furthermore, the population genetic effects of larger structural mutations must be considered at two levels; the effects of the mutated region as a whole and the effects of the loci within. To understand the evolutionary importance of various mutation types we must examine these effects.

We propose that to understand the contributions of different types of mutations to evolution, we must merge the rich history of population, quantitative and evolutionary genetics [1,3] with molecular genetics and genomics to connect mutations with evolutionary outcomes (Figure 1).

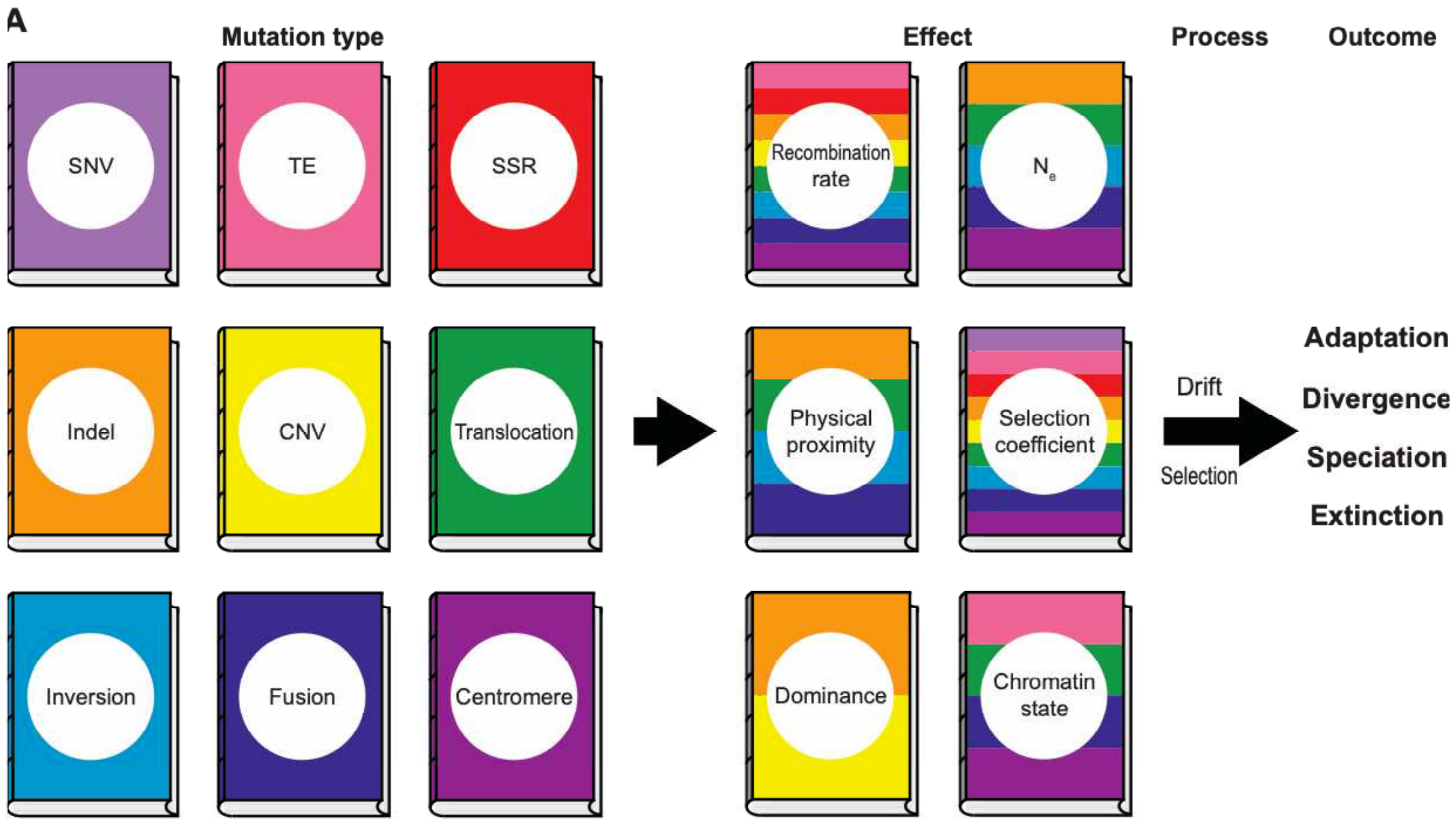

**B**

| | Recombination rate | $N_e$ | Physical proximity | Selection coefficient | Dominance | Chromatin state |
|---|---|---|---|---|---|---|
| SNV | - | - | - | ↓↑- | - | - |
| TE insertion | ↓↑ | - | - | ↓↑- | - | ↑ |
| SSR/microsatellite | ↑ | - | - | ↓↑- | - | - |
| Insertion/deletion | ↓ | ↓ | ↓↑ | ↓↑- | ↑ | - |
| CNV/duplication | ↓ | - | - | ↓↑- | ↓ | - |
| Translocation | ↓ | ↓ | ↓↑ | ↓↑- | - | ↓↑ |
| Inversion | ↓ | ↓ | ↓↑ | ↓↑- | - | - |
| Fusion | ↓ | ↓ | ↑ | ↓↑- | - | ↑ |
| Centromere shift | ↓↑ | ↓ | - | ↓↑- | - | ↓↑ |

**Figure 1 (key figure)** A. From mutation type to evolutionary outcome. Colors match mutation type here and in Figure 2. B. Interactions between mutation type and population genetic and genomic effects. Up and down arrows indicate an increase (up) or decrease (down) while a dash indicates no effect (known or unknown). Grey indicates that the relationship is variable and the mutation may have different effects depending on other factors. Indels and inversions are assumed to be large enough to affect pairing at meiosis. Smaller indels and inversions are expected to behave similarly to SNPs.

Here, we begin this process by reviewing the occurrence rates and known population genetic and genomic effects of different mutation types, and, using recombination rate as an example, suggest how to quantify the effect of the different mutation types (Box 1). We then discuss how these effects will influence evolutionary processes and finally how we may connect mutation type to evolutionary outcomes.

## Mutation Rates

Mutation rate is an essential parameter when investigating the evolutionary impact of different types of mutations. It can be measured indirectly, by comparing polymorphism data within and between closely related species, or directly, by comparing the number of mutations in gametes, zygotes or offspring [4]. Using these methods, estimates of the mutation rate vary greatly across taxa and mutation types (Figure 2). However, the range of estimated mutation rates for many different mutation types overlap. In addition, different mutations vary with respect to their detectability (for example, **SNPs** are much more likely to be detected with short-read sequencing data than larger insertions or deletions [5]). Note that we do not include fissions due to lack of information nor whole genome duplications as this have been covered extensively in other reviews [6]. More empirical data is sorely needed to be able to ascertain how the *de novo* mutation rate varies among mutation types and taxa.

## Population genetic and genomic effects of mutations

### Physical Linkage (physical proximity)

Large structural variants have the potential to affect physical proximity of multiple loci, either bringing previously unlinked loci into linkage or breaking apart pre-existing links. These effects can involve loci located on the same chromosome or on different chromosomes.

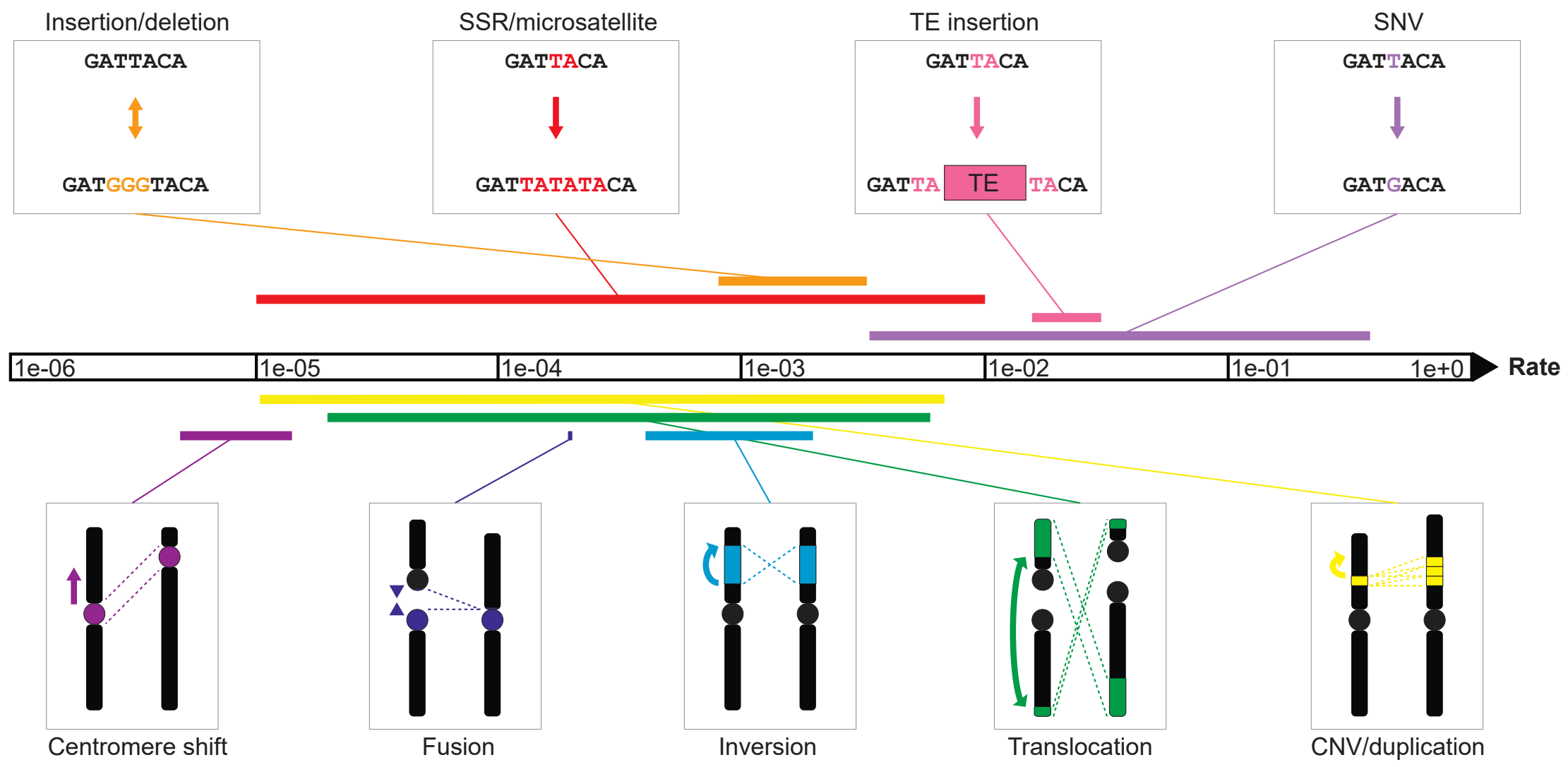

**Figure 2** - Overview of mutation types and mutation rates. Mutation rates are per genome per generation and are taken from [7–28]. Please note that this is not an exhaustive overview and that actual ranges shoud be larger.

Only **fusions** and **translocations** affect loci on different chromosomes and both drastically change physical linkage patterns, bringing previously completely (physically) unlinked loci into linkage. Additionally, translocations, unlike fusions, will also break physical linkage between the loci on either side of the breakpoints, moving them to separate chromosomes.

Both **inversions** and large **indels** will affect physical linkage between loci on the same chromosome. While large insertions will decrease proximity between loci either side, large deletions can bring loci closer together. By changing the position of loci in the genome, inversions affect physical linkage relationships between loci inside and outside the inverted region.

By changing linkage between loci, mutations may also change the regulatory environment of a gene, potentially altering gene expression patterns. However, this may not be common; artificial inversions in *Drosophila* did not lead to changes in gene expression, suggesting that such changes observed in natural populations are due to linked genetic variation [29]. Changes in methylation or chromatin state also alter the regulatory environment (see below).

## Dominance

Dominance is a complex parameter best examined on a case-by-case basis. However, there are a few trends that have been noted between mutation types and dominance. For example, as long as an insertion contains a single copy gene (i.e. a gene that is not present elsewhere in the genome) mutations in that gene will be dominantly expressed in the heterozygous state. **Copy number variants** (**CNV**s) themselves will alter the penetrance of a dominant mutation. For example, duplications of the recessive allele may nullify the dominant mutation or compensate identical alleles with low gene expression level [24].

The dominance effects of inversions depend on multiple factors. For example, an inversion might be underdominant if crossovers (COs) in the inversion region lead to unbalanced gametes (see below). On the other hand, recessive deleterious alleles can accumulate within both the standard and the inverted arrangement, generating (associative) overdominance at the level of the inversion because recessive alleles are shielded in inversion heterokaryotypes [30].

## Methylation and Chromatin State

The DNA methylation and chromatin state of a region may have strong implications on the regulatory environment of the genes present as well as on the rate of recombination. In general, increasing DNA methylation or heterochromatin will decrease recombination (crossovers are less likely in highly heterochromatic regions [31]) and gene expression although this is not a steadfast rule.

**Transposable element (TEs)** are in a constant arms race with the host and genomes have evolved multitudes of sequence-specific mechanisms for TE silencing via DNA methylation and repressive histone marks (e.g., H3K9me2 and H3K9me3) [32,33]. Notably, this change in methylation and chromatin state need not be restricted to a new TE insertion itself but can also spread into adjacent genomic regions, e.g., up to 20 kb away from TE insertions in *Drosophila melanogaster* [34].

Centromeres contain both centromeric chromatin (characterized by the CENP-A histone which is the foundation for the kinetochore) and repressive histone marks [35]. We thus expect that a **centromere shift** along a chromosome will not only reduce the recombination rate in the new pericentromeric region as discussed below, but also spread DNA methylation and repressive chromatin marks through increased accumulation of TEs due to the reduced recombination. This generates a positive feedback loop between recombination rate, new TE insertions, and chromatin state as previously proposed for regions of low recombination in general [36].

Translocations and fusions may also change methylation and chromatin state. For example, a study on humans found multiple differentially methylated positions with respect to a translocation, 93% of which mapped to the translocation breakpoints [37]. In mice with chromosomal fusions, the area of the chromosome with repressive histone marks was greatly expanded [38]. Research on large structural variants and methylation/chromatin state is in its infancy and more studies looking at non-model organisms outside the context of disease are sorely needed.

## Recombination

Local recombination rate can be affected by mutations in different ways. First, the actual rate of recombination may change. Second, recombination may proceed normally but crossovers and segregation problems may lead to the creation of unbalanced gametes that lead to inviable offspring, reducing the effective recombination rate. In the following, we review how different mutations alter recombination rate. We also address this question more quantitatively in Box 1. While the molecular processes underlying recombination are not understood in detail, and may vary between taxa, Box 1 represents a first attempt towards a more quantitative comparison between different mutation types.

### Changes in the rate of recombination

In eukaryotes, double strand breaks form during the pairing of the homologs in meiosis and are repaired via two pathways. (1) A crossover event (CO), the outcome of which is visible as a chiasma later in meiosis or (2) the break is repaired as a non-crossover (NCO) event. Gene conversion (GC) can occur in both pathways [39]. Different mutation types change the rate, distribution of recombination events, or pathway taken.

Several types of mutations can affect the alignment and pairing of homologs at the beginning of the recombination process. In inversion heterozygotes, proper synapsis in the inverted region and subsequent crossing over are slightly reduced [40]. A large heterozygous indel will generally form “unpaired DNA loops” preventing COs [41]. CNVs can also affect recombination in heterozygotes due to differences in chromosome length, effectively reducing recombination by inhibiting proper pairing [42]. Recombination may even be affected outside of the mutated area. For example, COs were suppressed in the regions around large artificial insertions in *C. elegans* [43].

Other mutations change the distribution of COs. The presence of fusions changes the rates and distribution of chiasmata in both homo- and heterozygotes in a range of mammals [44]. For example, in mice (*Mus musculus domesticus*), the number of chiasma correlates negatively with the number of fusions but the distribution of the chiasmata along the chromosomal arm varies between homozygotes and heterozygotes [38,45].

Centromeres and their surrounding pericentromeric regions generally reduce recombination [46]. As a result, centromere shifts reduce recombination in a new genomic region. Conversely, the region of the former inactivated centromere would then be free of centromere-associated recombination reduction. To our knowledge, these aspects of centromere evolution have yet to be appreciated in an evolutionary context.

Other mutations change the outcome of double-strand breaks and/or the GC rate. In inverted regions, double-strand breaks are more likely to be resolved as NCOs; however, the GC rate is unchanged [47,48]. Conversely, heterozygous translocations do not affect the ratio of COs to NCOs but reduce the GC rate [49].

Several other mutations can change the recombination landscape on a smaller scale. For example, TE insertion polymorphisms can actively increase or decrease recombination rates depending on the genomic context and type of TE [36]. Some TE insertions may attract repressive histone marks, thus locally decreasing recombination [33]. On the other hand, some TEs may contain sequence motifs that turn new insertions into recombination hotspots [36]. It is thus advisable to separately analyze TE insertion polymorphisms by chromatin context and TE type to account for the complex interplay of recombination-increasing and recombination-decreasing effects of TEs when investigating their effects on recombination.

Similarly, **simple sequence repeats (SSRs)** can act as recombination hotspots [50,51], especially for tri- and penta- nucleotide microsatellites [50]. This could be linked to the instability during crossing over due to length variation between sister chromatids [52]. Some studies also suggest that different microsatellite motifs have varying effects on recombination rate [50,51,53].

Indels may change chromosome length and therefore may alter the frequency of recombination between loci located either side of the indel [54]. This effect may become non-negligible if indels are large or numerous in a given genomic region.

Recombination produces unbalanced gametes

Recombination may occur normally but lead to the creation of unbalanced gametes. When inversions are heterozygous the inverted region can either pair **homosynaptically** or **heterosynaptically** [55]. Crossing over can only occur in homosynaptically paired regions and single crossovers in the inverted region will lead to gametes with unbalanced chromosomes (with potentially large duplications and deletions) in inversion heterozygotes [56]. Both types of pairing will decrease recombination overall but heterosynaptic pairing reduces the production of unbalanced gametes [55].

Alternatively, recombination can proceed normally and create balanced products but these products may fail to segregate properly. In translocation heterozygotes, the four involved chromosomes form a quadrivalent structure during meiosis. Segregation from this structure can lead to the creation of aneuploid gametes with a rate of 18% to more than 80% [57,58]. Similarly, nondisjunction rates in fusion heterozygotes may be elevated, ranging from 1.2% to 30% depending on the system [44].

## Effective Population Size

**Effective population size ($N_e$)** usually reflects the process of drift, which can be affected indirectly by mutations through their effect on recombination rate and their selection coefficient. Changes in recombination and selection (leading to selective sweeps or background selection) may affect local $N_e$ and sometimes the global $N_e$ as well [52,53].

## Box 1 - Structural variants as local recombination modifiers

Different types of mutations can affect the transfer of genetic material between homologs. To quantify this effect, we derive the probability, $P(x_1, x_2)$, that two loci at position $x_1$and $x_2$ (with $x_1 < x_2$), initially on the same homolog, are separated during meiosis. We present here only approximations obtained when the rate of double strand break (DSB) is sufficiently small (see Supplement for detailed expressions).
In the absence of a structural variant, the probability that two loci are separated by recombination is given by:

$$P_{rec}(x_1, x_2) \cong \beta_{DSB}\,(\lambda\, \phi_{GC} + \,(x_2 - x_1)\phi_{CO})$$

, with $\phi_{GC}$ and $\phi_{CO}$ as the probabilities that a DSB leads respectively to gene conversion (GC) and a crossover and $\lambda$ the length of a GC tract ($x_2 < x_1 + \lambda$). The first term corresponds to one locus being transferred by GC and the second to a crossover between the two focal loci.

Insertion/Deletion:
Recombination only happens in the ancestral (deletion) or derived (insertion) homozygote (its frequency denoted $f_{AA}$). The two loci are separated with probability:

$$P_{Indel}(x_1, x_2) = f_{AA}\, P_{rec}(x_1, x_2)$$

Inversion:
Single crossovers occurring within the inversion breakpoints in heterozygotes form gametes with unbalanced chromosomes, leading to inviable zygotes. Therefore, heterozygotes are underdominant and recombination only happens through GC or double crossovers. The probability that two loci in the inverted region are separated is given by (assuming $\beta_{DSB} \ll\; f_{Aa} - 0.5$):

$$P_{inv}(x) \cong \left(\frac{(x_2 - x_1)(1 - 2\, f_{Aa})\phi_{CO}}{1 - f_{Aa}} + \lambda\, \phi_{GC}\right)\beta_{DSB}$$

The first term corresponds to a recombination event happening between the focal loci in the homozygotes, whose frequency is increased due to underdominant heterozygotes. The second term corresponds to GC and remains unaffected. Double crossovers do not play a significant role under those conditions.

Fusion:
For chromosomal fusions, homologs in heterozygotes may fail to segregate properly (with probability $\beta_{NDJ}$), producing unbalanced gametes and reducing the contribution of heterozygotes to the next generation. In addition, the chance of a crossover decreases if at least one fused chromosome is involved. The two loci are separated with probability:

$$P_{fus}(x_1, x_2) \cong (\mathrm{S}_1(f_{Aa}, f_{AA}(\,(x_2 - x_1)\phi_{CO} + \lambda\phi_{GC})\beta_{DSB}$$

The contribution of crossovers is reduced by a factor $\mathrm{S}_1(f_{Aa}, f_{AA}($, which depends on the genotypes frequencies and captures both selection against the heterozygote and the reduced crossover probability when at least one fused chromosome is involved.

Translocation:
Similarly, homologs in heterozygotes may fail to segregate properly, producing unbalanced gametes and reducing the contribution of heterozygotes to the next generation. The GC rate in heterozygotes is also reduced. The two loci are separated with probability:

$$P_{trans}(x_1, x_2) \cong ((x_2 - x_1)\phi_{CO} + S_2(f_{Aa}(\lambda\phi_{GC})\beta_{DSB}$$

The contribution of gene conversion is reduced by a factor $S_2(f_{Aa})$, which depends on the frequency of the heterozygotes and captures both the effect of selection against, and the reduction of gene conversion within, heterozygotes.

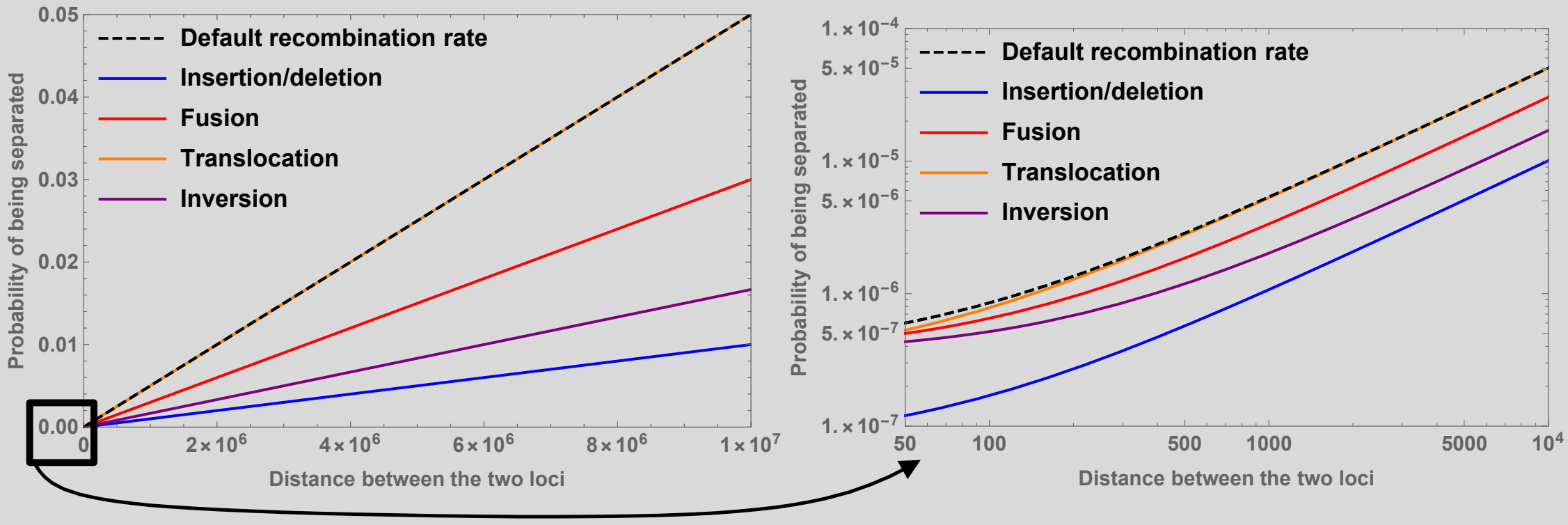

Figure Box 1. Probability that two loci on the same structural variant are separated due to recombination as a function of the distance between the two loci. Parameters: $\beta_{DSB} = 10^{-8}$; $\phi_{CO} = 0.5$; $\phi_{GC} = 0.7$; $\lambda = 50$; $f_{Aa} = 0.4$; $f_{AA} = 0.2$. The factors $S_1$ and $S_2$ are calculated using the expressions given in the Supplement; here $S_1 = -0.6$ and $S_2 = -0.8$.

Recombination and therefore $N_e$ can be reduced by a variety of structural variants. For example, loci within an indel will experience this reduction in $N_e$ twice, once due to the reduction in recombination (see Box 1) and once due to lower number of copies of the indel region. The indel as a locus, if neutral, will not be affected by either of these processes. Similarly, recombination between different arrangements of a polymorphic inversion is lowered and the arrangements can be viewed as two smaller and partially isolated populations [55,56]. Translocations and fusions will experience a similar effect.

Changes in fitness due to mutations can also lead reduction in $N_e$. For example, transposable elements are often considered weakly deleterious [42,54]. Translocations or chromosome fusions lead to high rates of non-disjunction and subsequent negative selection against heterozygotes.

Alternatively, centromere shifts may be under positive selection if they distort their segregation during female meiosis of heterozygotes, a process known as centromere drive [59].

## Selection Coefficient

All of the population genetic and genomic effects summarized above will affect the evolutionary processes: selection and drift (summarized in Table 1). This is quantified in the selection coefficient which measures differences in relative fitness, encompassing many population genetic effects. The selection coefficient thus bridges mutation and evolutionary process. While the selection coefficient is essential to determine/predict evolutionary outcomes, its estimation is highly complex, due to the large number of levels at which it can be affected.

All mutation types can be neutral or under positive/negative selection. The selection coefficient of a mutation depends on a multitude of factors including (1) the genomic context, i.e. whether it alters coding, regulatory, or intergenic regions, (2) whether it leads to sequence changes or positional shifts and (3) the selective environment (both extrinsic and intrinsic) where the change occurs [8,20,32,36,46–52,60,61]. Furthermore, duplicated regions, such as CNVs, have additional effects as they may free up selective constraints and can lead to the emergence of new gene functions [62]. Classic evolutionary examples involve a number of different mutation types. For example, a deletion modifying a regulatory region for the *Pitx-1*gene in sticklebacks is positively selected in freshwater environments [63] and a TE insertion in the first intron of the gene *cortex* controls the iconic industrial melanism of peppered moth [64].

For certain variants, selection can act at multiple levels. (i) The mutations themselves may be under negative selection, e.g. if they disrupt functional regions at breakpoints [65], or decrease fertility (see *Recombination*, above) [57,58]. (ii) Large variants may also experience selection based on their allelic content, e.g. inversions may be overdominant because recessive deleterious alleles are shielded in heterozygotes [66]. (iii) Large variants can also alter the efficacy of selection within the mutated region by modifying the recombination rate and local $N_e$. For example, inversions may be indirectly selected because they reduce recombination between multiple beneficial alleles located in the same arrangement (e.g. locally adapted alleles under gene flow [67]) (iv) Finally, the effects of certain mutations can spread outside of the mutated region. For example, many TEs encode genes that promote their replication independently, potentially rewiring the regulation of nearby genes [68].

The **distribution of fitness effects (DFE)** describes the relative frequency of different mutations and their fitness effects, basically summarizing the interaction of the mutation with drift and selection [69,70]. Most studies have estimated the DFE of SNPs or other **single nucleotide variants (SNVs)** and have found a bi- or multi-modal distribution, with beneficial mutations at low frequency, although the exact shape of distributions vary [69–71]. However, the DFE of other mutation types may have different properties (but see [72]). For example, a study in *E.coli* [73] showed a long flat deleterious tail and a high peak surrounding the neutral region for TE mutations. Most CNVs are expected to be found at the extremes of the distribution with either largely beneficial or deleterious effects [28]. Estimating the DFE for more mutation types and in different environments is a crucial step moving forward.

**Table 1** - Interactions between population genetic effects and evolutionary processes

| | Natural Selection | Drift |
|---|---|---|
| **Reduced Recombination rate** | ↕ Reduced/increased effect of selection depending on whether both focal and linked alleles are selected in the same direction (stronger Hill-Robertson effect).<br><br>↓ Potentially reduces the speed of adaptation<br><br>↑ Facilitates divergence in presence of divergent selection and linkage of co-adapted alleles<br><br>↑ Facilitates adaptation in presence of epistatic selection and linkage of co-adapted alleles | ↑ Increases the impact of stochastic effects and decreases $N_e$ when together with selection (similar to drift). |
| **Reduced $N_e$** | ↓ Reduced contribution of selection (compared to drift) | ↑ Increased contribution of drift (relative to selection) |
| **Physical Linkage** | ↕ Increase/decrease of the effect of selection, dependent on which loci are located in the new linkage background | ↕ Increased/reduced stochasticity, depending on which loci are located in the new linkage background |
| Dominance Coefficient | ↑ An increase in h leads to directional or negative selection being more efficient.<br><br>↑ In the case of overdominance via beneficial alleles, an increase in h leads to selection being more efficient, with an increase in frequency of the heterozygote<br><br>↑ In the case of frequency-dependent selection, an increase in h will help spread the rare derived allele due to penetrance in the heterozygote. Similarly, a decrease in h will help spread the rare ancestral allele. | ↓ An increase in h leads to selection being more efficient. Therefore the relative contribution of drift compared to selection is reduced. |
| Selection Coefficient | ↑ Increase in magnitude of the selection coefficient increases the effect of directional or negative selection<br><br>↑ Increase in the magnitude of selection coefficient in heterozygotes will speed up the decrease of homozygotes frequency under overdominance<br><br>↑ Under negative frequency dependent selection, larger selection coefficients will lead to a faster increase of the rare allele (assuming it has an increased selection coefficient). | ↓ Increase in the magnitude of the selection coefficient may make it less likely that the mutation is fixed or lost due to genetic drift |

## Connecting mutational effects to evolution

A variety of bottom up and top down approaches, including DFE estimation, can be used to bridge the gap between mutation type and evolutionary processes. Starting from the bottom, more empirical studies examining the population genetic effects of mutations are sorely needed for better characterization of mutation effects and to determine how these effects vary across taxa. The information from these studies can also be incorporated into theoretical models (see Box 1 for an example) to generate new predictions. From the top down, population genomic studies can be strengthened by examining multiple types of mutations together. This will require collecting different types of genomic data sets (e.g. short- and long-read re-sequencing and mapping crosses) from the same population and developing detection pipelines targeted at different mutation types. Feeding this information to population genetic models would allow for indirect estimation of DFEs for all mutational types. Along with better estimates of mutation rates, these approaches will enable the quantification of the evolutionary significance of different mutation types.

Quantifying the relationships between mutation type, population genetic effects and the major evolutionary processes, selection and genetic drift (summarized in Table 1), allows us to draw connections to evolutionary outcomes. For example, speciation requires the build-up of linkage disequilibrium both within and between different reproductive isolation barriers [74]. Mutations that reduce recombination should aid speciation with gene flow by protecting this nascent linkage disequilibrium. We thus predict that mutations such as inversions, indels, TEs, and centromere shifts might be major drivers of speciation events [75]. A critical next step would be testing some of these hypotheses, for example using a meta-analysis (see Outstanding Questions).

## Concluding remarks and future perspectives

Our framework highlights the fact that mutations may affect evolution in several ways and that many different mutation types have similar population genetic effects (Box 1, Figure 1). We suggest that shifting the focus away from the structure or size of a mutation to its effects facilitates directly linking mutations with evolutionary outcomes.

To better understand the link between mutation types and evolutionary outcomes, comparable measurements of mutation rates as well as the population genetic and genomic effects of different mutation types are urgently needed. In particular, much of the current data comes from either model systems or research on diseases. Characterizing different mutation types in non-model organisms and natural populations is critical for the future. Understanding the evolutionary significance of different mutations will require unboxing them and viewing their effects in a larger population genetic context.

## Outstanding Questions

**What are the rates of occurrence for different mutation types and does this vary within or between species?**

*Comprehensive studies across taxa using both natural populations and traditional mutation accumulation lines will help us better estimate occurrence rates. This information is invaluable for determining the contribution of different mutations for different evolutionary outcomes.*

**What are the population genetic and genomic effects of different mutation types? Are these consistent across taxa?**

*For example, fine-scale recombination studies in multiple taxa are necessary to determine the effects of different mutations on both crossing over and gene conversion rates. Studies using techniques such as ChIP-seq and ATAC-seq will allow us to explore the chromatin state of non-model organisms and to understand how different mutations may alter this.*

**Are certain mutation types (e.g. inversions) more important than others for specific evolutionary outcomes (e.g. adaptation), and why?**

*Genomic studies that analyze a variety of mutation types rather than 1-2 will provide key insights into the relative roles of different mutation types in different outcomes (e.g. adaptation). This type of data will be key for allowing more targeted analyses to be done in the future. When such studies have accumulated, a meta-analysis testing for a relationship between mutation type and evolutionary outcome is the next step towards answering this question more generally.*

**How can we compare the contributions of different mutation types in practice?**

*Expanding both empirical and theoretical studies to include a wide variety of mutation types will help move the field forward. Theoretical studies can incorporate the global effects of different mutations on parameters (such as recombination, see Box 1) to better understand how these aspects contribute to evolutionary outcomes. Empirical studies can work to better estimate*

*mutation rates and understand the population genetic and genomic effects of mutations (see above). Incorporating multiple types of mutations in population genomic studies will be critical for linking mutation type with evolutionary outcome. Feeding this information to population genetic models would allow for indirect estimation of DFEs for all mutational types, a crucial step for moving forward.*

---

## Glossary

**Single Nucleotide Polymorphism (SNP) -** A single base pair substitution in a specific position in the genome.

**Copy Number Variant (CNV) -** A DNA segment of at least one kb that is present at a variable copy number.

**Centromere shift -** Repositioning of the centromere along the chromosome.

**Distribution of Fitness Effects (DFE) -** Describes the proportion of new mutations that are beneficial, deleterious or neutral in a specific environment.

**Transposable Element (TE) -** A selfish genetic element propagating via copy-and-paste or cut-and-paste.

**Simple Sequence Repeat (SSR) -** Tandem repeats of 2-6 bp motifs.

**Hill-Robertson effect -** describes selection having a reduced effect when selected sites are in tight linkage with other selected sites.

**Translocation** (**also called balanced or reciprocal translocation)** -Two pieces of non-homologous chromosomes that have broken off and been switched.

**Fusion (also called Robertsonian fusion, Robertsonian translocation,** and **centric fusion) -** Two acrocentric chromosomes (where the centromere is located near the end of the chromosome) that have experienced breaks at or near the centromere and then fused creating a metacentric chromosome.

**Homosynaptic pairing** - When the two homologs correctly synapse during prophase 1.

**Heterosynaptic pairing** - When non-homologous (heterologous) synapsis occurs during prophase 1.

**Effective population size, $N_e$** - The equivalent population size of a WrightFisher population that will generate population genetic statistics closest to the ones of the focal population.

**Effective recombination rate (*sensu* Golding)** - The equivalent recombination rate in a Wright Fisher population that will generate the same linkage disequilibrium patterns as those found in the focal population.

**Indel** - Small genetic variant from 10 to 10 000 bp that can be either inserted or deleted from the genome

**Inversion** - A segment of the genome that is rotated 180 degrees.

**Single Nucleotide Variant (SNV)** - A change to a single base pair. Either an exchange, insertion or a deletion.

## Acknowledgements

We thank to Roger Butlin, Kerstin Johannesson, Valentina Peona, Rike Stelkens, Julie Blommaert, Nick Barton, and João Alpedrinha for helpful comments that improved the manuscript. The authors acknowledge funding from the Swedish Research Council Formas (2017-01597 to AS), the Swedish Research Council Vetenskapsrådet (2016-05139 to AS, 2019-04452 to TS) and from the European Research Council (ERC) under the European Union's Horizon 2020 research and innovation programme (grant agreement No 757451 to TS). ELB was funded by a Carl Tryggers grant awarded to TS. AMW was funded by the European Union's Horizon 2020 research and innovation programme under the Marie Sklodowska-Curie grant agreement No 797747. IF was funded by a Junior Researcher contract from FCT (CEECIND/02616/2018).

## References

1. Futuyma, D.J. (1986). Reflections on reflections: ecology and evolutionary biology. J. Hist. Biol. 19, 303–312.

2. Mérot, C., Oomen, R.A., Tigano, A., and Wellenreuther, M. (2020). A Roadmap for Understanding the Evolutionary Significance of Structural Genomic Variation. Trends Ecol. Evol. 35, 561–572.

3. Charlesworth, B. (2010). Elements of Evolutionary Genetics (Roberts Publishers).

4. Fu, Y.-X., and Huai, H. (2003). Estimating mutation rate: how to count mutations? Genetics 164, 797–805.

5. Ho, S.S., Urban, A.E., and Mills, R.E. (2020). Structural variation in the sequencing era. Nat. Rev. Genet. 21, 171–189.

6. Fox, D.T., Soltis, D.E., Soltis, P.S., Ashman, T.-L., and Van de Peer, Y. (2020). Polyploidy: A biological force from cells to ecosystems. Trends Cell Biol. 30, 688–694.

7. Maeda, T., Ohno, M., Matsunobu, A., Yoshihara, K., and Yabe, N. (1991). A cytogenetic survey of 14,835 consecutive liveborns. Jinrui Idengaku Zasshi 36, 117–129.

8. Ducos, A., Berland, H.-M., Bonnet, N., Calgaro, A., Billoux, S., Mary, N., Garnier-Bonnet, A., Darré, R., and Pinton, A. (2007). Chromosomal control of pig populations in France: 2002-2006 survey. Genet. Sel. Evol. 39, 583–597.

9. Yamaguchi, O., and Mukai, T. (1974). Variation of spontaneous occurrence rates of chromosomal aberrations in the second chromosomes of *Drosophila melanogaster*. Genetics 78, 1209–1221.

10. Vendrell-Mir, P., Barteri, F., Merenciano, M., González, J., Casacuberta, J.M., and Castanera, R. (2019). A benchmark of transposon insertion detection tools using real data. Mobile DNA 10.

11. Goerner-Potvin, P., and Bourque, G. (2018). Computational tools to unmask transposable elements. Nat. Rev. Genet. 19, 688–704.

12. Feusier, J., Scott Watkins, W., Thomas, J., Farrell, A., Witherspoon, D.J., Baird, L., Ha, H., Xing, J., and Jorde, L.B. (2019). Pedigree-based estimation of human mobile element retrotransposition rates. Genome Research 29, 1567–1577.

13. Sung, W., Ackerman, M.S., Dillon, M.M., Platt, T.G., Fuqua, C., Cooper, V.S., and Lynch, M. (2016). Evolution of the insertion-deletion mutation rate across the tree of life. G3 6, 2583–2591.

14. Ramu, A., Noordam, M.J., Schwartz, R.S., Wuster, A., Hurles, M.E., Cartwright, R.A., and Conrad, D.F. (2013). DeNovoGear: *de novo* indel and point mutation discovery and phasing. Nat. Methods 10, 985–987.

15. Farlow, A., Long, H., Arnoux, S., Sung, W., Doak, T.G., Nordborg, M., and Lynch, M. (2015). The spontaneous mutation rate in the fission yeast *Schizosaccharomyces pombe*. Genetics 201, 737–744.

16. Schrider, D.R., Houle, D., Lynch, M., and Hahn, M.W. (2013). Rates and genomic consequences of spontaneous mutational events in *Drosophila melanogaster*. Genetics 194, 937–954.

17. Jarne, P., and Lagoda, P.J. (1996). Microsatellites, from molecules to populations and back. Trends Ecol. Evol. 11, 424–429.

18. Thuillet, A.-C., Bru, D., David, J., Roumet, P., Santoni, S., Sourdille, P., and Bataillon, T. (2002). Direct estimation of mutation rate for 10 microsatellite loci in durum wheat, *Triticum turgidum* (L.) Thell. ssp *durum* desf. Mol. Biol. Evol. 19, 122–125.

19. Marriage, T.N., Hudman, S., Mort, M.E., Orive, M.E., Shaw, R.G., and Kelly, J.K. (2009). Direct estimation of the mutation rate at dinucleotide microsatellite loci in *Arabidopsis thaliana* (Brassicaceae). Heredity 103, 310–317.

20. Gemayel, R., Vinces, M.D., Legendre, M., and Verstrepen, K.J. (2010). Variable tandem repeats accelerate evolution of coding and regulatory sequences. Annu. Rev. Genet. 44, 445–477.

21. Rocchi, M., Archidiacono, N., Schempp, W., Capozzi, O., and Stanyon, R. (2012). Centromere repositioning in mammals. Heredity 108, 59–67.

22. Marshall, O.J., Chueh, A.C., Wong, L.H., and Choo, K.H.A. (2008). Neocentromeres: New insights into centromere structure, disease development, and karyotype evolution. Am. J. Hum. Genet. 82, 261–282.

23. Fu, W., Zhang, F., Wang, Y., Gu, X., and Jin, L. (2010). Identification of copy number variation hotspots in human populations. Am. J. Hum. Genet. 87, 494–504.

24. Beckmann, J.S., Estivill, X., and Antonarakis, S.E. (2007). Copy number variants and genetic traits: Closer to the resolution of phenotypic to genotypic variability. Nat. Rev. Genet. 8, 639–646.

25. Brumfield, R.T., Beerli, P., Nickerson, D.A., and Edwards, S.V. (2003). The utility of single nucleotide polymorphisms in inferences of population history. Trends Ecol. Evol. 18, 249–256.

26. Weng, M.-L., Becker, C., Hildebrandt, J., Neumann, M., Rutter, M.T., Shaw, R.G., Weigel, D., and Fenster, C.B. (2019). Fine-grained analysis of spontaneous mutation spectrum and frequency in *Arabidopsis thaliana*. Genetics 211, 703–714.

27. Ossowski, S., Schneeberger, K., Lucas-Lledó, J.I., Warthmann, N., Clark, R.M., Shaw,

R.G., Weigel, D., and Lynch, M. (2010). The rate and molecular spectrum of spontaneous mutations in *Arabidopsis thaliana*. Science 327, 92–94.

28. Katju, V., and Bergthorsson, U. (2013). Copy-number changes in evolution: rates, fitness effects and adaptive significance. Front. Genet. 4, 273.

29. Said, I., Byrne, A., Serrano, V., Cardeno, C., Vollmers, C., and Corbett-Detig, R. (2018). Linked genetic variation and not genome structure causes widespread differential expression associated with chromosomal inversions. Proceedings of the National Academy of Sciences 115, 5492–5497.

30. Ohta, T. (1971). Associative overdominance caused by linked detrimental mutations. Genet. Res. 18, 277–286.

31. Henderson, I.R. (2012). Control of meiotic recombination frequency in plant genomes. Curr. Opin. Plant Biol. 15, 556–561.

32. Hollister, J.D., and Gaut, B.S. (2009). Epigenetic silencing of transposable elements: a trade-off between reduced transposition and deleterious effects on neighboring gene expression. Genome Res. 19, 1419–1428.

33. Choi, J.Y., and Lee, Y.C.G. (2020). Double-edged sword: The evolutionary consequences of the epigenetic silencing of transposable elements. PLoS Genet. 16, e1008872.

34. Lee, Y.C.G., and Karpen, G.H. (2017). Pervasive epigenetic effects of *Drosophila* euchromatic transposable elements impact their evolution. eLife 6.

35. Sullivan, B.A., and Karpen, G.H. (2004). Centromeric chromatin exhibits a histone modification pattern that is distinct from both euchromatin and heterochromatin. Nat. Struct. Mol. Biol. 11, 1076–1083.

36. Kent, T.V., Uzunović, J., and Wright, S.I. (2017). Coevolution between transposable elements and recombination. Philos. Trans. R. Soc. Lond. B Biol. Sci. 372.

37. McCartney, D.L., Walker, R.M., Morris, S.W., Anderson, S.M., Duff, B.J., Marioni, R.E., Millar, J.K., McCarthy, S.E., Ryan, N.M., Lawrie, S.M., et al. (2018). Altered DNA methylation associated with a translocation linked to major mental illness. NPJ Schizophr 4, 5.

38. Capilla, L., Medarde, N., Alemany-Schmidt, A., Oliver-Bonet, M., Ventura, J., and Ruiz-Herrera, A. (2014). Genetic recombination variation in wild Robertsonian mice: On the role of chromosomal fusions and Prdm9 allelic background. Proc. Biol. Sci. 281.

39. Hunter, N. (2015). Meiotic recombination: The essence of heredity. Cold Spring Harb. Perspect. Biol. 7.

40. Gong, W.J., McKim, K.S., and Hawley, R.S. (2005). All paired up with no place to go: pairing, synapsis, and DSB formation in a balancer heterozygote. PLoS Genet. 1, e67.

41. Poorman, P.A., Moses, M.J., Russell, L.B., and Cacheiro, N.L. (1981). Synaptonemal complex analysis of mouse chromosomal rearrangements. I. Cytogenetic observations on a tandem duplication. Chromosoma 81, 507–518.

42. Sjödin, P., and Jakobsson, M. (2012). Population genetic nature of copy number variation. Methods in Molecular Biology, 209–223.

43. Hammarlund, M., Davis, M.W., Nguyen, H., Dayton, D., and Jorgensen, E.M. (2005). Heterozygous insertions alter crossover distribution but allow crossover interference in *Caenorhabditis elegans*. Genetics 171, 1047–1056.

44. Dobigny, G., Britton-Davidian, J., and Robinson, T.J. (2017). Chromosomal polymorphism in mammals: An evolutionary perspective. Biol. Rev. Camb. Philos. Soc. 92, 1–21.

45. Bidau, C.J., Giménez, M.D., Palmer, C.L., and Searle, J.B. (2001). The effects of Robertsonian fusions on chiasma frequency and distribution in the house mouse (*Mus musculus domesticus*) from a hybrid zone in northern Scotland. Heredity 87, 305–313.

46. Stapley, J., Feulner, P.G.D., Johnston, S.E., Santure, A.W., and Smadja, C.M. (2017). Variation in recombination frequency and distribution across eukaryotes: Patterns and processes. Philos. Trans. R. Soc. Lond. B Biol. Sci. 372.

47. Korunes, K.L., and Noor, M.A.F. (2019). Pervasive gene conversion in chromosomal inversion heterozygotes. Mol. Ecol. 28, 1302–1315.

48. Crown, K.N., Miller, D.E., Sekelsky, J., and Hawley, R.S. (2018). Local inversion heterozygosity alters recombination throughout the genome. Curr. Biol. 28, 2984–2990.e3.

49. Sherizen, D., Jang, J.K., Bhagat, R., Kato, N., and McKim, K.S. (2005). Meiotic recombination in *Drosophila* females depends on chromosome continuity between genetically defined boundaries. Genetics 169, 767–781.

50. Guo, W.-J., Ling, J., and Li, P. (2009). Consensus features of microsatellite distribution: Microsatellite contents are universally correlated with recombination rates and are preferentially depressed by centromeres in multicellular eukaryotic genomes. Genomics 93, 323–331.

51. Brandström, M., Bagshaw, A.T., Gemmell, N.J., and Ellegren, H. (2008). The relationship between microsatellite polymorphism and recombination hot spots in the human genome. Mol. Biol. Evol. 25, 2579–2587.

52. Kayser, M., Vowles, E.J., Kappei, D., and Amos, W. (2006). Microsatellite length differences between humans and chimpanzees at autosomal Loci are not found at equivalent haploid Y chromosomal Loci. Genetics 173, 2179–2186.

53. Myers, S., Bottolo, L., Freeman, C., McVean, G., and Donnelly, P. (2005). A fine-scale map of recombination rates and hotspots across the human genome. Science 310, 321–324.

54. Smukowski, C.S., and Noor, M.A.F. (2011). Recombination rate variation in closely related species. Heredity 107, 496–508.

55. Torgasheva, A.A., and Borodin, P.M. (2010). Synapsis and recombination in inversion heterozygotes. Biochem. Soc. Trans. 38, 1676–1680.

56. Rieseberg, L.H. (2001). Chromosomal rearrangements and speciation. Trends in Ecology & Evolution 16, 351–358.

57. Morel, F., Douet-Guilbert, N., Le Bris, M.-J., Herry, A., Amice, V., Amice, J., and De Braekeleer, M. (2004). Meiotic segregation of translocations during male gametogenesis. Int. J. Androl. 27, 200–212.

58. Talukdar, D. (2010). Reciprocal translocations in grass pea (*Lathyrus sativus L.*): pattern of transmission, detection of multiple interchanges and their independence. J. Hered. 101, 169–176.

59. Malik, H.S. (2009). The centromere-drive hypothesis: a simple basis for centromere complexity. Prog. Mol. Subcell. Biol. 48, 33–52.

60. Weissensteiner, M.H., Bunikis, I., Catalán, A., Francoijs, K.-J., Knief, U., Heim, W., Peona, V., Pophaly, S.D., Sedlazeck, F.J., Suh, A., et al. (2020). Discovery and population genomics of structural variation in a songbird genus. Nature Communications 11.

61. Flynn, J.M., Rossouw, A., Cote-Hammarlof, P., Fragata, I., Mavor, D., Hollins, C., Bank, C., and Bolon, D.N.A. (2020). Comprehensive fitness maps of Hsp90 show widespread environmental dependence. eLife 9.

62. Ohno, S. (2013). Evolution by Gene Duplication (Springer Science & Business Media).

63. Chan, Y.F., Marks, M.E., Jones, F.C., Villarreal, G., Jr, Shapiro, M.D., Brady, S.D., Southwick, A.M., Absher, D.M., Grimwood, J., Schmutz, J., et al. (2010). Adaptive evolution of pelvic reduction in sticklebacks by recurrent deletion of a Pitx1 enhancer. Science 327, 302–305.

64. Van't Hof, A.E., Campagne, P., Rigden, D.J., Yung, C.J., Lingley, J., Quail, M.A., Hall, N., Darby, A.C., and Saccheri, I.J. (2016). The industrial melanism mutation in British peppered moths is a transposable element. Nature 534, 102–105.

65. Kirkpatrick, M. (2010). How and why chromosome inversions evolve. PLoS Biol. 8.

66. Ota, T. (1971). Associative overdominance caused by linked detrimental mutations. Genet. Res. 18, 277–286.

67. Kirkpatrick, M., and Barton, N. (2006). Chromosome inversions, local adaptation and speciation. Genetics 173, 419–434.

68. Chuong, E.B., Elde, N.C., and Feschotte, C. (2017). Regulatory activities of transposable

elements: from conflicts to benefits. Nat. Rev. Genet. 18, 71–86.

69. Eyre-Walker, A., and Keightley, P.D. (2007). The distribution of fitness effects of new mutations. Nat. Rev. Genet. 8, 610–618.

70. Keightley, P.D., and Eyre-Walker, A. (2010). What can we learn about the distribution of fitness effects of new mutations from DNA sequence data? Philos. Trans. R. Soc. Lond. B Biol. Sci. 365, 1187–1193.

71. Bataillon, T., and Bailey, S.F. (2014). Effects of new mutations on fitness: Insights from models and data. Annals of the New York Academy of Sciences 1320, 76–92.

72. Barton, H.J., and Zeng, K. (2018). New methods for inferring the distribution of fitness effects for INDELs and SNPs. Mol. Biol. Evol. 35, 1536–1546.

73. Elena, S.F., Ekunwe, L., Hajela, N., Oden, S.A., and Lenski, R.E. (1998). Distribution of fitness effects caused by random insertion mutations in *Escherichia coli*. Genetica 102-103, 349–358.

74. Butlin, R.K., and Smadja, C.M. (2018). Coupling, Reinforcement, and Speciation. Am. Nat. 191, 155–172.

75. Fuller, Z.L., Leonard, C.J., Young, R.E., Schaeffer, S.W., and Phadnis, N. (2018). Ancestral polymorphisms explain the role of chromosomal inversions in speciation. PLoS Genet. 14, e1007526.

**Appendix**

In this appendix, we derive the probability, $P(x_1, x_2)$, that two loci at position $x_1$and $x_2$ (with $x_1 < x_2$), initially on the same homolog, are separated during meiosis. We derive this quantity for loci affected by various structural variants. However, we first present this probability in the absence of any structural variants.

In the rest of the appendix, we will use the following convention to indicate the position of a recombination event: a crossover or a gene conversion event begins at position $i$, indicates that the break happens between position $i$ and position $i + 1$.

<u>Default case:</u>

We initially consider the effect of gene conversion alone. We define $\beta_{DSB}$ the rate of double strand break (DSB) per base pair, $\phi_{GC}$ the probability that the DSB generates a GC event and $\lambda$ the length of a gene conversion event. We initially assume that the loci at position $x_1$ and $x_2$ are sufficiently apart ($x_2 - x_1 > \lambda$). An allele at position $x_1$ is transferred to the homolog if a gene conversion event is initiated at position $i$, with this initiating bp, $i$, being at most $\lambda - 1$ base pair upstream from $x_1$. In that case, allele $x_1$ will be transferred to the homolog and separated from $x_2$ with probability:

$$\frac{\beta_{DSB}\phi_{GC}}{2}\left(1 - \frac{\beta_{DSB}\phi_{GC}}{2}\right)^{\lambda}$$

The first factor corresponds to the probability that a DSB happens at position $i$ and generates a GC event. The factor two is due to the unidirectionality of gene conversion, i.e. the sequence of the focal chromosome is copied to the homolog, so $x_1$and $x_2$ are separated only when the focal chromosome is "receiving" the sequence. The second factor corresponds to the probability of not having a crossover that may move allele $x_2$, from the homolog to the focal chromosome.

To obtain the probability that the allele at position $x_1$, we need to sum this probability over all possible positions of $i$ that may affect $x_1$, ranging from $x_1 - \lambda$, where $x_1$ is the last base pair affected by gene conversion, to $x_1 - 1$, where the gene conversion event starting just before $x_1$. The same process happens if $x_2$ is the transferred allele leading to:

$$P_{GC}(x_1, x_2) = \sum_{i=x_1-\lambda}^{x_1-1} \frac{\beta_{DSB}\phi_{GC}}{2}\left(1 - \frac{\beta_{DSB}\phi_{GC}}{2}\right)^{\lambda} + \sum_{i=x_2-\lambda}^{x_2-1} \frac{\beta_{DSB}\phi_{GC}}{2}\left(1 - \frac{\beta_{DSB}\phi_{GC}}{2}\right)^{\lambda}$$

$$= \lambda\, \beta_{DSB}\phi_{GC}\left(1 - \frac{\beta_{DSB}\phi_{GC}}{2}\right)^{\lambda}$$

We are now considering loci that are relatively close from each other ($x_2 - x_1 < \lambda$). First, we consider that the GC event is affecting the locus at position $x_1$. Where the GC event may start is limited by the position $x_2$. Indeed, the GC event can start up to $\lambda$ base pair upstream from $x_1$; however the last position it can take is $\lambda + 1$ base pairs upstream from $x_2$. For a GC event affecting the locus at position $x_2$, it can only start after $x_1$. Therefore, we obtain the following expression for the two loci to be separated by GC, when they are sufficiently close:

$$P_{GC}(x_1,x_2) = \sum_{i=x_1-\lambda}^{x_2-\lambda-1} \frac{\beta_{DSB}\phi_{GC}}{2}\left(1-\frac{\beta_{DSB}\phi_{GC}}{2}\right)^{\lambda} + \sum_{i=x_1}^{x_2-1} \frac{\beta_{DSB}\phi_{GC}}{2}\left(1-\frac{\beta_{DSB}\phi_{GC}}{2}\right)^{\lambda}$$

$$= (x_2-x_1)\,\beta_{DSB}\phi_{GC}\left(1-\frac{\beta_{DSB}\phi_{GC}}{2}\right)^{\lambda}$$

Alternatively, a double stand break can also generate a crossover with probability $\phi_{CO}$. The two loci are separated by a single crossover at position $i$ (with $x_1 \le i < x_2$) with probability:

$$\beta_{DSB}\phi_{CO}(1-\beta_{DSB}\,\phi_{CO})^{x_2-x_1-1}$$

The first term corresponds to the crossover happening at position $i$, with no other crossovers happening between $x_1$ and $x_2-1$ (a CO happening at position $x_2$ does not affect $x_2$). We then sum this probability over all possible positions for $i$, ranging from $x_1$ to $x_2-1$:

$$P_{1CO}(x_1,x_2) = \sum_{i=x_1}^{x_2-1} \beta_{DSB}\phi_{CO}(1-\beta_{DSB}\,\phi_{CO})^{x_2-x_1-1}$$

$$= (x_2-x_1)\beta_{DSB}\phi_{CO}(1-\beta_{DSB}\,\phi_{CO})^{x_2-x_1-1}$$

In the absence of a structural variant, the two loci are separated by recombination (ignoring double crossovers and other higher order recombination events) is given by probability:

$$P_{rec}(x_1,x_2) = (1-\beta_{DSB}\phi_{CO})^{x_2-x_1}P_{GC}(x_1,x_2) + \left(1-\frac{\beta_{DSB}\phi_{GC}}{2}\right)^{2\lambda} P_{1CO}(x_1,x_2)$$

The first term corresponds to one locus being transferred by GC, without interference from a crossover; the second term to one locus being transferred by a crossover, without interference from GC at either loci.

This expression simplifies to:

$$P_{rec}(x_1,x_2) \cong \beta_{DSB}\,(\lambda\,\phi_{GC} + (x_2-x_1)\phi_{CO}) + o(\beta_{\mathrm{DSB}}^2)$$

if double strand breaks are rare enough and the two loci are sufficiently apart ($x_2 - x_1 > \lambda$). If the two loci are close from each other ($x_2 - x_1 < \lambda$), $P_{rec}(x_1,x_2)$ simplifies to:

$$P_{rec}(x_1,x_2) \cong (x_2-x_1)(\phi_{GC}+\phi_{CO})\beta_{DSB} + o(\beta_{DSB}^2)$$

In the rest of the appendix, we will calculate the probability that two loci, found in a structural variant, are separated. The ancestral state is denoted $a$ and the derived allele $A$.

Insertion/deletion:

Recombination only happens in the ancestral homozygotes for a deletion and the derived homozygotes for an insertion. For simplicity, we define the frequency of the ancestral homozygote $f_{AA}$ and the derived homozygote $f_{aa}$. Recombination inside the structural variant is therefore a function of the frequency of the mutant (insertion) or ancestral (deletion) genotype.

$$P_{In}(x_1, x_2) = f_{AA}\, P_{rec}(x_1, x_2)$$

$$P_{Del}(x_1, x_2) = f_{aa}\, P_{rec}(x_1, x_2)$$

Inversion:

For inversions, recombination is suppressed between the standard and inverted arrangement. More precisely single crossovers within the inversion breakpoints (denoted here $y_0$ and $y_1$) in heterozygotes form gametes with unbalanced chromosomes and therefore reduces the contribution of heterozygotes to the next generation. The frequency of heteroyzygotes after selection, $f'_{Aa}$, is therefore given by (assuming soft selection):

$$f'_{Aa} = \frac{f_{Aa}(1 - P_{CO}(y_0, y_1))}{1 - f_{Aa}(1 - P_{CO}(y_0, y_1))}$$

Since single crossover do not contribute to recombination, transfer between arrangements only happens through GC or double crossovers ($P_{2CO}$), with both double strand breaks (denoted $i$ and $j$) happening within the inversion.

The probability that two loci, in the inverted region, are separated by a double crossover is:

$$P_{2CO}(x_1, x_2) = \sum_{i=y_0-1}^{x_1-1} \beta_{DSB}\phi_{CO}(1 - \beta_{DSB}\phi_{CO})^{x_1-y_0} \sum_{j=x_1}^{x_2-1} \beta_{DSB}\phi_{CO}(1 - \beta_{DSB}\phi_{CO})^{x_2-x_1-1} + \sum_{i=x_1}^{x_2-1} \beta_{DSB}\phi_{CO}(1 - \beta_{DSB}\phi_{CO})^{x_2-x_1-1} \sum_{j=x_2}^{y_1} \beta_{DSB}\phi_{CO}(1 - \beta_{DSB}\phi_{CO})^{y_1-x_2}$$

The first term corresponds to the locus at position $x_1$ being transferred to the homolog ($\beta_{DSB}\phi_{CO}(1 - \beta_{DSB}\phi_{CO})^{x_1-y_0}$), summed over all possible positions where the first double strand break, $i$, can happen, ranging from $y_0 - 1$ (the double strand break can happen in front of the inversion itself) to $x_1 - 1$(the last position before $x_1$) .The second term ($\beta_{DSB}\phi_{CO}(1 - \beta_{DSB}\phi_{CO})^{x_2-x_1-1}$) corresponds to the second double strand break, $j$, happening after $x_1$ but before position $x_2$. Alternatively, the first double strand break can happen between $x_1$and $x_2 - 1$, and the second double strand break between $x_2$ and the end of the inversion $y_1$(the double strand break can happen on the inversion breakpoint itself).

After simplification, we obtain:

$$P_{2CO}(x_1, x_2) = \frac{(x_2 - x_1)\beta_{DSB}^2\phi_{CO}^2}{1 - \beta_{DSB}\phi_{CO}}\left((x_1 - y_0 + 1)(1 - \beta_{DSB}\phi_{CO})^{x_2-y_0} + (1 + y_1 - x_2)(1 - \beta_{DSB}\phi_{CO})^{y_1-x_1}\right)$$

This expression further simplifies to:

$$P_{2CO}(x_1, x_2) \cong (x_2 - x_1)(2 + x_1 - x_2 - y_0 + y_1)\phi_{CO}^2\beta_{DSB}^2 + o(\beta_{DSB}^3)$$

There is no first order term, meaning that double crossovers can be safely ignored if the rate of double breakpoints is small enough (as long as $x_2 - x_1 \ll \beta_{DSB}$ and $y_1 - y_0 \ll \beta_{DSB}$).

The probability that the two alleles, at position $x_1$and $x_1$ are separated is given by:

$$P_{Inv}(x_1, x_2) = P_{rec}(x_1, x_2)(1 - f'_{Aa}) + f'_{Aa}\left(P_{GC}(x_1, x_2)(1 - \beta_{DSB}\phi_{CO})^{y_1 - y_0} + P_{2CO}(x_1, x_2)\left(1 - \frac{\beta_{DSB}\phi_{GC}}{2}\right)^{2\lambda}\right)$$

The first term corresponds to recombination happening unchanged in the two homozygotes (after selection). The second term corresponds to what happens in heterozygotes (after selection), with either the alleles being separated by gene conversion, $P_{GC}(x_1, x_2)$, if no crossover happens over the whole inversion $(1 - \beta_{DSB}\phi_{CO})^{y_1 - y_0}$ or the alleles are separated by a double crossover $P_{2CO}(x_1, x_2)$ and the two loci are not affected by gene conversion $\left(1 - \frac{\beta_{DSB}\phi_{GC}}{2}\right)^{2\lambda}$. If $\beta_{DSB}$ is small enough, the two loci are sufficiently far apart and the frequency of the heterozygotes differs from $\frac{1}{2}$, $P_{inv}(x)$ simplifies to

$$P_{inv}(x) \cong \left(\frac{(x_2 - x_1)(1 - 2 f_{Aa})\phi_{CO}}{1 - f_{Aa}} + \lambda\,\phi_{GC}\right)\beta_{DSB} + o(\beta_{DSB}^2)$$

If the two loci are close enough, we obtain:

$$P_{inv}(x) \cong (x_2 - x_1)\left(\frac{(1 - 2 f_{Aa})\phi_{CO}}{1 - f_{Aa}} + \phi_{GC}\right)\beta_{DSB} + o(\beta_{DSB}^2)$$

Fusion:

In the case of chromosomal fusions or translocations, the homologs in heterozygotes may fail to segregate properly (with a probability $\beta_{NDJ}$), producing gametes with unbalanced chromosomes, leading to a decrease in the contribution of heterozygotes to the next generation:

$$f'_{Aa} = \frac{f_{Aa}(1 - \beta_{NDJ})}{1 - f_{Aa}\,\beta_{NDJ}} \text{ and } f'_{AA} = \frac{f_{AA}}{1 - f_{Aa}\,\beta_{NDJ}}$$

In addition, the chance of a crossover is reduced by a factor $h$, as long as at least one fused chromosome is involved (for the heterozygote and the derived homozygote). The probability that 2 loci are separated is given by:

$$P_{fus}(x_1, x_2) = P_{rec}(x_1, x_2)(1 - f'_{Aa} - f'_{AA}) + \left((1 - \beta_{DSB}\phi_{CO})^{x_2 - x_1} P_{GC}(x_1, x_2) + (1 - h)\left(1 - \frac{\beta_{DSB}\phi_{GC}}{2}\right)^{2\lambda} P_{1CO}(x)\right)(f'_{Aa} + f'_{AA})$$

The first term corresponds to recombination happening normally in the ancestral homozygote (after selection). The second term corresponds to recombination happening either in the heterozygote or derived ancestral homozygote, with the reduction by a factor $h$ in the probability of having a crossover. After simplifying, we obtain:

$$P_{fus}(x_1, x_2) = P_{rec}(x_1, x_2) - \frac{(h\,P_{1CO}(x_1, x_2))(f_{AA} + f_{Aa}(1 - \beta_{NDJ}))\left(1 - \frac{\beta_{DSB}\phi_{GC}}{2}\right)^{2\lambda}}{1 - f_{Aa}\,\beta_{NDJ}}$$

If double strand breaks are sufficiently rare and the two loci sufficiently apart, $P_{fus}(x_1, x_2)$ simplifies to:

$$P_{fus}(x_1, x_2) \cong \left(\frac{(x_2 - x_1)\big(1 - (f_{Aa} + f_{AA})h + f_{Aa}(-1 + h)\beta_{NDJ}\big)\phi_{CO}}{1 - f_{Aa}\beta_{NDJ}} + \lambda\phi_{GC}\right)\beta_{DSB} + o(\beta_{DSB}^2)$$

In the main text, we use the variable $S_1$, to simplify the expression:

$$S_1 = \frac{\big(1 - (f_{Aa} + f_{AA})h + f_{Aa}(-1 + h)\beta_{NDJ}\big)}{1 - f_{Aa}\beta_{NDJ}}$$

If the two loci are close from each other, we obtain:

$$P_{fus}(x_1, x_2) \cong (x_2 - x_1)\left(\frac{\big(1 - (f_{Aa} + f_{AA})h + f_{Aa}(-1 + h)\beta_{NDJ}\big)\phi_{CO}}{1 - f_{Aa}\beta_{NDJ}} + \phi_{GC}\right)\beta_{DSB} + o(\beta_{DSB}^2)$$

<u>Translocation:</u>

In the case of translocations, gene conversion is impeded in heterozygotes and is reduced by a factor $c$. The probability that 2 alleles, at position $x_1$ and $x_2$, are separated by recombination is given by:

$$P_{trans}(x_1, x_2) = P_{rec}(x_1, x_2)(1 - f'_{Aa}) + f'_{Aa}\left(\left(1 - (1 - c)\frac{\beta_{DSB}\phi_{GC}}{2}\right)^{2\lambda} P_{1CO}(x_1, x_2) + (1 - c)(1 - \beta_{DSB}\,\phi_{CO})^{x_2 - x_1} P_{GC}(x_1, x_2)\right)$$

The first term corresponds to recombination happening without change in the two homozygotes (after selection). The second term focuses on the heterozygotes (after selection), with the two loci separated by a crossover, $P_{1CO}(x_1, x_2)$, without the interference of gene conversion $\left(1 - (1 - c)\frac{\beta_{DSB}\phi_{GC}}{2}\right)^{2\lambda}$ or by gene conversion, reduced by the factor $c$, $(1 - c)\, P_{GC}(x_1, x_2)$ and in the absence of gene crossover $(1 - \beta_{DSB}\,\phi_{CO})^{x_2 - x_1}$. After simplification, we obtain:

$$P_{trans}(x_1, x_2) = P_{rec}(x)(1 + \frac{f_{AA}}{1 - f_{Aa}\,\beta_{NDJ}}\left(-1 + \left(\frac{2 + (c - 1)\beta_{DSB}\,\phi_{GC}}{2 - \beta_{DSB}\,\phi_{GC}}\right)^{2\lambda}\right) + \frac{f_{AA}}{1 - f_{Aa}\,\beta_{NDJ}}(1 - \beta_{DSB}\,\phi_{CO})^{x_2 - x_1} P_{GC}(x_1, x_2)\left(1 - c - \left(\frac{2 + (c - 1)\beta_{DSB}\,\phi_{GC}}{2 - \beta_{DSB}\,\phi_{GC}}\right)^{2\lambda}\right)$$

This expression further simplifies, if $\beta_{DSB}$ is small enough, to:

$$P_{trans}(x_1, x_2) \cong \left( (x_2 - x_1)\phi_{CO} + \frac{\left(-1 + f_{Aa}\left(c + \beta_{NDJ}(1-c)\right)\right)\lambda\phi_{GC}}{-1 + f_{Aa}\beta_{NDJ}} \right)\beta_{DSB} + o(\beta_{DSB}^2)$$

.

In the main text, we use the variable $S_2$, to simplify the expression:

$$S_2 = \frac{\left(-1 + f_{Aa}\left(c + \beta_{NDJ}(1-c)\right)\right)\lambda\phi_{GC}}{-1 + f_{Aa}\beta_{NDJ}}$$

If the two loci are close from each other, we obtain:

$$P_{trans}(x_1, x_2) \cong (x_2 - x_1)\left( \phi_{CO} + \frac{\left(-1 + f_{Aa}\left(c + \beta_{NDJ}(1-c)\right)\right)\phi_{GC}}{-1 + f_{Aa}\beta_{NDJ}} \right)\beta_{DSB} + o(\beta_{DSB}^2)$$

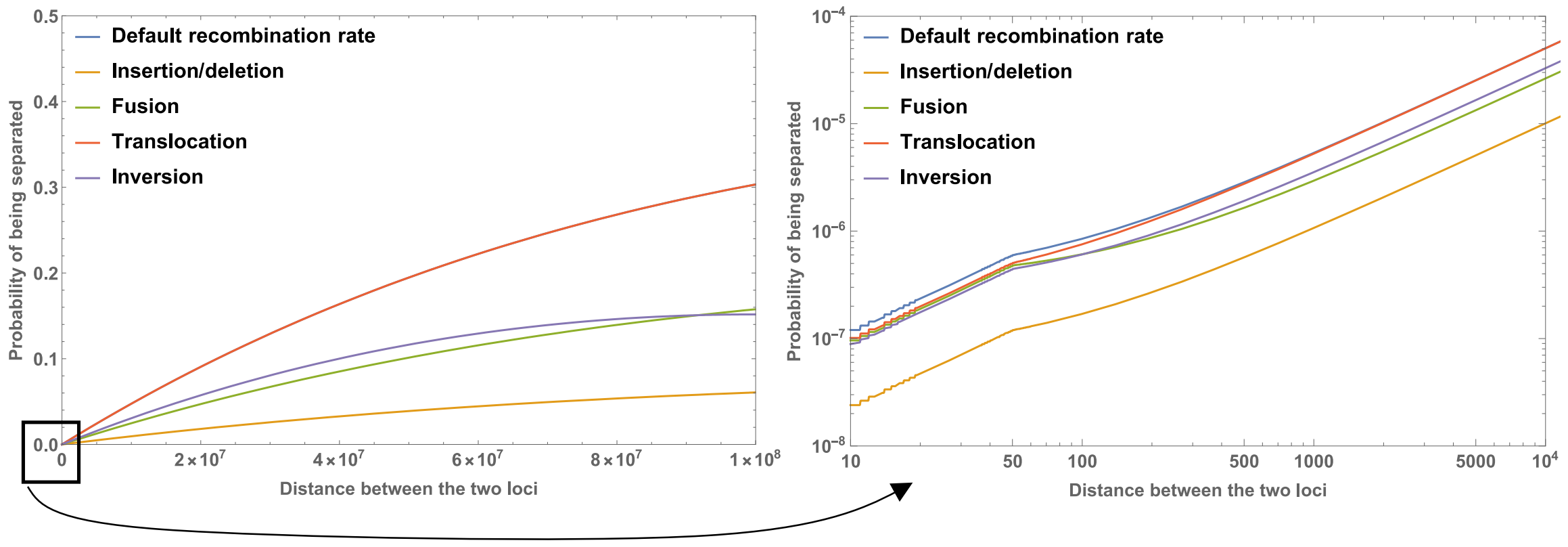

Figure. Probability that two loci on the same structural variant are separated due to recombination as a function of the distance between the two loci, using the expressions derived above (solid lines) and the approximations given here and presented in the main manuscript (dotted lines). This corresponds to the figure presented in Box 1 in the main manuscript. Parameters: $\beta_{DSB} = 10^{-8}$; $\phi_{CO} = 0.5$; $\phi_{GC} = 0.7$; $\lambda = 50$; $f_{Aa} = 0.5$; $f_{AA} = 0.2$; $h = 0.8$; $\beta_{NDJ} = 0.5$; $y_0 = 1$; $y_1 = 10^8$; $c = 0.8$; $x_1 = 10^5$

# Structural variants as recombination modifier

## Default case:

### Gene conversion:

Probability that two alleles at positions x1 and x2 are separated by gene conversion is given by the function PGC:

```
PGC[βDSB_, ϕGC_, λ_, x1_, x2_] :=
 If[x2 - x1 > λ, 2 * Sum[βDSB * ϕGC / 2 , {i, x1 - λ, x1 - 1}] * (1 - βDSB ϕGC / 2) ^λ,
  Sum[βDSB * ϕGC / 2 , {i, x1 - λ, x2 - λ - 1}] (1 - βDSB ϕGC / 2) ^λ +
   (1 - βDSB ϕGC / 2) ^λ Sum[βDSB * ϕGC / 2 , {i, x1, x2 - 1}]]
```

This expression simplifies to:

```
PGC[βDSB, ϕGC, λ, x1, x2] // FullSimplify
```

$$\mathrm{If}\left[x2 > x1 + \lambda,\ 2\left(\sum_{i=x1-\lambda}^{x1-1} \frac{\beta DSB\ \phi GC}{2}\right)\left(1-\frac{\beta DSB\ \phi GC}{2}\right)^{\lambda},\ \left(\sum_{i=x1-\lambda}^{x2-\lambda-1} \frac{\beta DSB\ \phi GC}{2}\right)\left(1-\frac{\beta DSB\ \phi GC}{2}\right)^{\lambda} + \left(1-\frac{\beta DSB\ \phi GC}{2}\right)^{\lambda}\sum_{i=x1}^{x2-1}\frac{\beta DSB\ \phi GC}{2}\right]$$

$$\mathrm{FullSimplify}\left[2\left(\sum_{i=x1-\lambda}^{x1-1} \frac{\beta DSB\ \phi GC}{2}\right)\left(1-\frac{\beta DSB\ \phi GC}{2}\right)^{\lambda}\right]$$

$$\beta DSB\ \lambda\ \phi GC\left(1-\frac{\beta DSB\ \phi GC}{2}\right)^{\lambda}$$

$$\mathrm{FullSimplify}\left[\left(\sum_{i=x1-\lambda}^{x2-\lambda-1} \frac{\beta DSB\ \phi GC}{2}\right)\left(1-\frac{\beta DSB\ \phi GC}{2}\right)^{\lambda} + \left(1-\frac{\beta DSB\ \phi GC}{2}\right)^{\lambda}\sum_{i=x1}^{x2-1}\frac{\beta DSB\ \phi GC}{2}\right]$$

$$(-x1 + x2)\ \beta DSB\ \phi GC\left(1-\frac{\beta DSB\ \phi GC}{2}\right)^{\lambda}$$

If βDSB is sufficiently small, we obtain the following approximation for PGC:

```
Series[PGC[βDSB, ϕGC, λ, x1, x2], {βDSB, 0, 1}]
```

$$\begin{cases} \lambda\ \phi GC\ \beta DSB + O[\beta DSB]^2 & x1 - x2 + \lambda < 0 \\ (-x1\ \phi GC + x2\ \phi GC)\ \beta DSB + O[\beta DSB]^2 & \mathrm{True} \end{cases}$$

### Single crossover

Probability that two alleles at positions x1 and x2 are separated by a single crossover is given by the function PCO:

```
PCO[βDSB_, ϕCO_, x1_, x2_] := Sum[βDSB ϕCO (1 - βDSB ϕCO) ^ (x2 - x1 - 1), {i, x1, x2 - 1}]
```

This expression simplifies to:

```
FullSimplify[PCO[βDSB, ϕCO, x1, x2]]
```

$(-x1 + x2)\ \beta DSB\ \phi CO\ (1 - \beta DSB\ \phi CO)^{-1-x1+x2}$

If βDSB is sufficiently small, we obtain the following approximation for PCO:

```
Series[(-x1 + x2) βDSB ϕCO (1 - βDSB ϕCO)^(-1-x1+x2), {βDSB, 0, 1}]
```

$(-x1\ \phi CO + x2\ \phi CO)\ \beta DSB + O[\beta DSB]^2$

## Recombination

From the two previous cases, we can now calculate the probability that two alleles at positions x1 and x2 are separated by recombination. It is given by the function Prec:

```
Prec[βDSB_, ϕCO_, ϕGC_, λ_, x1_, x2_] :=
  (1 - βDSB ϕCO) ^ (x2 - x1) PGC[βDSB, ϕGC, λ, x1, x2] +
   (1 - βDSB ϕGC / 2) ^ (2 λ) PCO[βDSB, ϕCO, x1, x2]
```

If βDSB is sufficiently small, we obtain the following approximation for Prec:

```
Series[Prec[βDSB, ϕCO, ϕGC, λ, x1, x2], {βDSB, 0, 1}] // FullSimplify
```

$$\begin{cases} -(x1 - x2)\ (\phi CO + \phi GC)\ \beta DSB + O[\beta DSB]^2 & x1 + \lambda \geq x2 \\ (-x1\ \phi CO + x2\ \phi CO + \lambda\ \phi GC)\ \beta DSB + O[\beta DSB]^2 & \text{True} \end{cases}$$

# Structural variants

In this section, we focus on the fate of alleles that are both captured by the structural variant.

## Indel

For indels, recombination only happens in the ancestral homozygotes (for deletion) or derived homozygotes (for insertion).

```
PIndel[βDSB_, ϕCO_, ϕGC_, λ_, fAA_, x1_, x2_] := (Prec[βDSB, ϕCO, ϕGC, λ, x1, x2]) fAA
```

## Inversion

In individuals heterozygous for the inversion, single crossovers generate unbalanced chromosomes and therefore the produced gametes lead to inviable zygote. It means that one needs to consider double crossovers when deriving the probability that the two alleles at x1 and x2 are separated.

```
PDCO[βDSB_, ϕCO_, y0_, y1_, x1_, x2_] :=
 Sum[βDSB ϕCO (1 - βDSB ϕCO ) ^ (x1 - y0) Sum[βDSB ϕCO (1 - βDSB ϕCO ) ^ (x2 - x1 - 1),
     {j, x1, x2 - 1}], {i, y0 - 1, x1 - 1}] + Sum[βDSB ϕCO (1 - βDSB ϕCO ) ^ (x2 - x1 - 1)
    Sum[βDSB ϕCO (1 - βDSB ϕCO ) ^ (y1 - x2), {i, x2, y1}], {i, x1, x2 - 1}]
```

It simplifies to:

```
FullSimplify[PDCO[βDSB, ϕCO, y0, y1, x1, x2]]
```

$$\frac{1}{1 - \beta DSB\ \phi CO} (-x1 + x2)\ \beta DSB^2\ \phi CO^2 \left( (1 + x1 - y0)\ (1 - \beta DSB\ \phi CO)^{x2-y0} + (1 - x2 + y1)\ (1 - \beta DSB\ \phi CO)^{-x1+y1} \right)$$

Due to the production of gametes with unbalanced chromosomes, the contribution of heterozygotes is reduced by a factor (1-PCO[βDSB,ϕCO,y0,y1])/(1-fAa(1-PCO[βDSB,ϕCO,y0,y1]). The probability that the two alleles at position x1 and x2 are separated by recombination is given by:

```
PInversion[βDSB_, ϕCO_, ϕGC_, λ_, fAa_, y0_, y1_, x1_, x2_] :=
 Prec[βDSB, ϕCO, ϕGC, λ, x1, x2]
   (1 - fAa (1 - PCO[βDSB, ϕCO, y0, y1]) / (1 - fAa (1 - PCO[βDSB, ϕCO, y0, y1]))) +
  fAa (1 - PCO[βDSB, ϕCO, y0, y1]) / (1 - fAa (1 - PCO[βDSB, ϕCO, y0, y1]))
   (PGC[βDSB, ϕGC, λ, x1, x2] (1 - βDSB ϕCO) ^ (y1 - y0) +
     (((-x1 + x2) βDSB^2 ϕCO^2 ((1 + x1 - y0) (1 - βDSB ϕCO)^(x2-y0) +
            (1 - x2 + y1) (1 - βDSB ϕCO)^(-x1+y1))) / (1 - βDSB ϕCO)) (1 - βDSB ϕGC / 2) ^ (2 λ))
```

If βDSB is sufficiently small, we obtain the following approximation for PInversion :

```
Series[PInversion[βDSB, ϕCO, ϕGC, λ, fAa, y0, y1, x1, x2], {βDSB, 0, 1}] // FullSimplify
```

$$\begin{cases} -\frac{(x1-x2)\ (\phi CO-2\ fAa\ \phi CO+\phi GC-fAa\ \phi GC)\ \beta DSB}{1-fAa}+O[\beta DSB]^2 & x1+\lambda\geq x2 \\ \left(-\frac{(-1+2\ fAa)\ (x1-x2)\ \phi CO}{-1+fAa}+\lambda\ \phi GC\right)\ \beta DSB+O[\beta DSB]^2 & \text{True} \end{cases}$$

We are providing the term of second order since the term of first order may be cancelled if fAa=0.5.

```
FullSimplify[
 Series[PInversionRestricted[βDSB, ϕCO, ϕGC, λ, fAa, y0, y1, x1, x2], {βDSB, 0, 2}]]
```

$$\begin{cases} -\frac{(x1-x2)\ (\phi CO-2\ fAa\ \phi CO+\phi GC-fAa\ \phi GC)\ \beta DSB}{1-fAa}+\frac{1}{2\ (-1+fAa)^2}\ (x1-x2)\ \Big(-2\ (1+x1-x2+ fAa\ (-1+(-2+fAa)\ x1+2\ x2-2\ y0+2\ y1-fAa\ (x2-y0+y1)))\ \phi CO^2- 2\ (-1+fAa)\ ((-1+2\ fAa)\ x1+x2+\lambda-fAa\ (2\ x2+y0-y1+2\ \lambda))\ \phi CO\ \phi GC+ (-1+fAa)^2\ \lambda\ \phi GC^2\Big)\ \beta DSB^2+O[\beta DSB]^3 & x1+\lambda\geq x2 \\ \left(-\frac{(-1+2\ fAa)\ (x1-x2)\ \phi CO}{-1+fAa}+\lambda\ \phi GC\right)\ \beta DSB-\frac{1}{2\ (-1+fAa)^2}\ \Big(2\ (x1-x2)\ (1+x1-x2+ fAa\ (-1+(-2+fAa)\ x1+2\ x2-2\ y0+2\ y1-fAa\ (x2-y0+y1)))\ \phi CO^2- 2\ (-1+fAa)\ ((-2+4\ fAa)\ x1+2\ x2+fAa\ (-4\ x2-y0+y1))\ \lambda\ \phi CO\ \phi GC+ (-1+fAa)^2\ \lambda^2\ \phi GC^2\Big)\ \beta DSB^2+O[\beta DSB]^3 & \text{True} \end{cases}$$

## Fusion

For fusion, heterozygotes and derived homozygotes may also produce unbalanced gametes if the homologs fail to separate properly. The probability that the two alleles at position x1 and x2 are separated by recombination is given by:

```
PFusion[βDSB_, ϕCO_, ϕGC_, λ_, fAA_, fAa_, h_, βNDJ_, x1_, x2_] :=
 Prec[βDSB, ϕCO, ϕGC, λ, x1, x2] * (1 - fAa (1 - βNDJ) / (1 - fAa βNDJ) - fAA / (1 - fAa βNDJ)) +
  (fAa (1 - βNDJ) / (1 - fAa βNDJ) + fAA / (1 - fAa βNDJ))
   ((1 - βDSB ϕCO) ^ (x2 - x1) PGC[βDSB, ϕGC, λ, x1, x2] +
     (1 - h) (1 - βDSB ϕGC / 2) ^ (2 λ) PCO[βDSB, ϕCO, x1, x2])
```

If βDSB is sufficiently small, we obtain the following approximation for PFusion :

```
Series[PFusion[βDSB, ϕCO, ϕGC, λ, fAA, fAa, h, βNDJ, x1, x2], {βDSB, 0, 1}] //
 FullSimplify
```

$$\begin{cases} \frac{1}{-1+fAa\ \beta NDJ}\ (x1-x2)\ ((1-(fAa+fAA)\ h+fAa\ (-1+h)\ \beta NDJ)\ \phi CO+\phi GC-fAa\ \beta NDJ\ \phi GC)\ \beta DSB+ O[\beta DSB]^2 & x1+\lambda\geq x2 \\ (((x1-x2)\ (1-(fAa+fAA)\ h+fAa\ (-1+h)\ \beta NDJ)\ \phi CO)\ /\ (-1+fAa\ \beta NDJ)+\lambda\ \phi GC)\ \beta DSB+O[\beta DSB]^2 & \text{True} \end{cases}$$

Since GC is not affected here, we replaced PGC using the definition of Prec to generate a clearer expression:

```
Prec * (1 - fAa (1 - βNDJ) / (1 - fAa βNDJ) - fAA / (1 - fAa βNDJ)) +
   (fAa (1 - βNDJ) / (1 - fAa βNDJ) + fAA / (1 - fAa βNDJ))
    ((1 - βDSB ϕCO) ^ (x2 - x1) PGC + (1 - h) (1 - βDSB ϕGC / 2) ^ (2 λ) PCO) /.
  {PGC -> (Prec - (1 - βDSB ϕGC / 2) ^ (2 λ) PCO) / ((1 - βDSB ϕCO) ^ (x2 - x1))} // FullSimplify
```

$$\text{Prec} + \frac{\text{h PCO } (\text{fAa} + \text{fAA} - \text{fAa } \beta\text{NDJ}) \left(1 - \frac{\beta\text{DSB } \phi\text{GC}}{2}\right)^{2\lambda}}{-1 + \text{fAa } \beta\text{NDJ}}$$

## Translocation

For translocation, heterozygotes may also produce unbalanced gametes if the homologs fail to separate properly. The probability that the two alleles at position x1 and x2 are separated by recombination is given by :

```
PTranslocation[βDSB_, ϕCO_, ϕGC_, λ_, fAa_, c_, βNDJ_, x1_, x2_] :=
 (Prec[βDSB, ϕCO, ϕGC, λ, x1, x2]) (1 - fAa (1 - βNDJ) / (1 - fAa βNDJ)) +
  fAa (1 - βNDJ) / (1 - fAa βNDJ) ((1 - (1 - c) βDSB ϕGC / 2) ^ (2 λ) PCO[βDSB, ϕCO, x1, x2] +
     (1 - c) (1 - βDSB ϕCO) ^ (x2 - x1) PGC[βDSB, ϕGC, λ, x1, x2])
```

If $\beta$DSB is sufficiently small, we obtain the following approximation for PTranslocation :

```
Series[PTranslocation[βDSB, ϕCO, ϕGC, λ, fAa, c, βNDJ, x1, x2], {βDSB, 0, 1}] //
 FullSimplify
```

$$\begin{cases} -\frac{1}{-1+\text{fAa } \beta\text{NDJ}} (\text{x1} - \text{x2}) \\ \quad ((-1 + \text{fAa } \beta\text{NDJ})\, \phi\text{CO} + (-1 + \text{fAa } (\text{c} + \beta\text{NDJ} - \text{c } \beta\text{NDJ}))\, \phi\text{GC})\, \beta\text{DSB} + O[\beta\text{DSB}]^2 & \text{x1} + \lambda \geq \text{x2} \\ \left(-\text{x1 } \phi\text{CO} + \text{x2 } \phi\text{CO} + \frac{(-1+\text{fAa } (\text{c}+\beta\text{NDJ}-\text{c } \beta\text{NDJ}))\, \lambda\, \phi\text{GC}}{-1+\text{fAa } \beta\text{NDJ}}\right) \beta\text{DSB} + O[\beta\text{DSB}]^2 & \text{True} \end{cases}$$

Since CO is not affected here, we replaced PCO using the definition of Prec to generate a clearer expression :

```
FullSimplify[
 Pr (1 - f) + f ((1 - (1 - c) β ϕGC / 2) ^ (2 λ) Pco + (1 - c) (1 - βDSB ϕCO) ^ (x2 - x1) Pgc) /.
  {Pco → (Pr - (1 - βDSB ϕCO) ^ (x2 - x1) Pgc) / (1 - β ϕGC / 2) ^ (2 λ)}]
```

$$\text{Pr} - \text{f Pr} + \text{f} \left(-(-1+\text{c}) \text{ Pgc } (1 - \beta\text{DSB } \phi\text{CO})^{-\text{x1}+\text{x2}} + \left(\text{Pr} - \text{Pgc } (1 - \beta\text{DSB } \phi\text{CO})^{-\text{x1}+\text{x2}}\right) (2 - \beta\, \phi\text{GC})^{-2\lambda} (2 + (-1+\text{c})\, \beta\, \phi\text{GC})^{2\lambda}\right)$$

```
CoefficientList[
  Pr - f Pr + f (- (-1 + c) Pgc (1 - βDSB ϕCO)^(-x1+x2) + (Pr - Pgc (1 - βDSB ϕCO)^(-x1+x2)) (2 - β ϕGC)^(-2 λ)
     (2 + (-1 + c) β ϕGC)^(2 λ)), Pr] // FullSimplify
```

$$\left\{\text{f Pgc } (1 - \beta\text{DSB } \phi\text{CO})^{-\text{x1}+\text{x2}} \left(1 - \text{c} - (2 - \beta\, \phi\text{GC})^{-2\lambda} (2 + (-1+\text{c})\, \beta\, \phi\text{GC})^{2\lambda}\right),\ 1 + \text{f} \left(-1 + (2 - \beta\, \phi\text{GC})^{-2\lambda} (2 + (-1+\text{c})\, \beta\, \phi\text{GC})^{2\lambda}\right)\right\}$$

## Figure

Illustration of the contribution of gene conversion, single and double crossovers

```
With[{βDSB = 10^-8, ϕCO = 0.5, ϕGC = 0.7, λ = 50, fAa = 0.4,
  fAA = 0.2, h = 0.8, βNDJ = .5, x1 = 100000, c = 0.8, y0 = 1, y1 = 10000000},
 LogLogPlot[{PGC[βDSB, ϕGC, λ, x1, x1 + d], PCO[βDSB, ϕCO, x1, x1 + d],
   (d βDSB^2 ϕCO^2 ((1 + x1 - y0) (1 - βDSB ϕCO)^(d+x1-y0) + (1 - d - x1 + y1) (1 - βDSB ϕCO)^(-x1+y1))) /
    (1 - βDSB ϕCO)}, {d, 10, 10^7}, PlotRange → All,
  PlotLegends → Placed[{"GC alone", "CO alone", "DCO alone"}, {Left, Center}],
  BaseStyle → {Thick, Bold, 14}, ImageSize → 600, Frame → True,
  FrameLabel → {"Distance between the two loci", "Probability of being separated"}]]
```

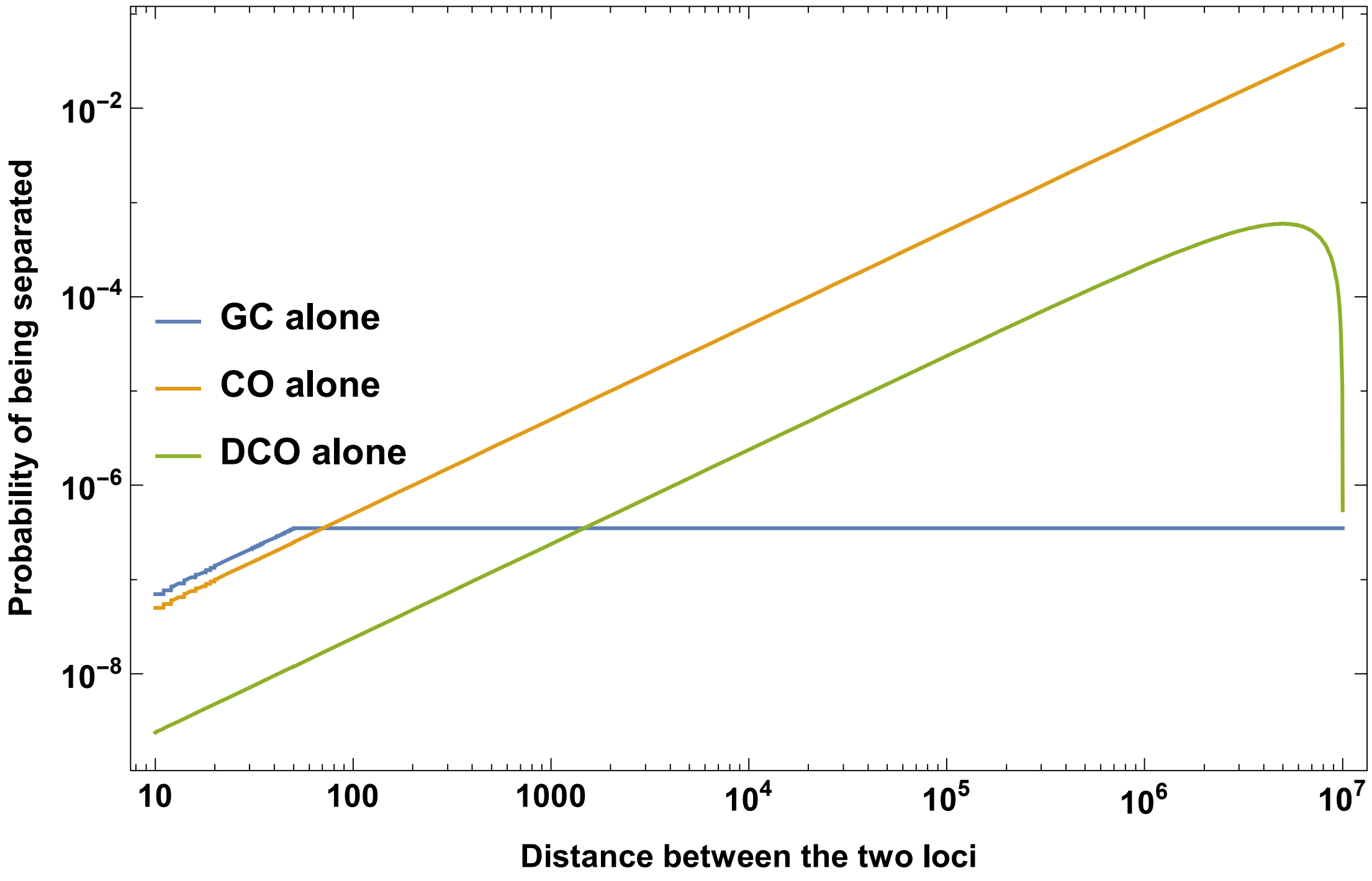

Illustration of the main expressions

```
With[{βDSB = 10^-8, ϕCO = 0.5, ϕGC = 0.7, λ = 50, fAa = 0.4,
  fAA = 0.2, h = 0.8, βNDJ = .5, y0 = 1, y1 = 10 000 000, x1 = 100 000, c = 0.8},
 Plot[{Prec[βDSB, ϕCO, ϕGC, λ, x1, x1 + d], PIndel[βDSB, ϕCO, ϕGC, λ, fAA, x1, x1 + d],
   PFusion[βDSB, ϕCO, ϕGC, λ, fAA, fAa, h, βNDJ, x1, x1 + d],
   PTranslocation[βDSB, ϕCO, ϕGC, λ, fAa, c, βNDJ, x1, x1 + d], PInversion[βDSB,
    ϕCO, ϕGC, λ, fAa, y0, y1, x1, x1 + d], Prec[βDSB, ϕCO, ϕGC, λ, x1, x1 + d]},
  {d, 10, y1 - x1}, PlotRange → {{10, 10^7}, {10^-8, 0.05}},
  PlotStyle → {{Dashed, Black}, Blue, Red, Orange, Purple, {Dashed, Black}},
  PlotLegends → Placed[{"Default recombination rate", "Insertion/deletion",
     "Fusion", "Translocation", "Inversion"}, {Left, Top}],
  BaseStyle → {Thick, Bold, 12}, ImageSize → 450, Frame → True,
  FrameLabel → {"Distance between the two loci", "Probability of being separated"}]]
```

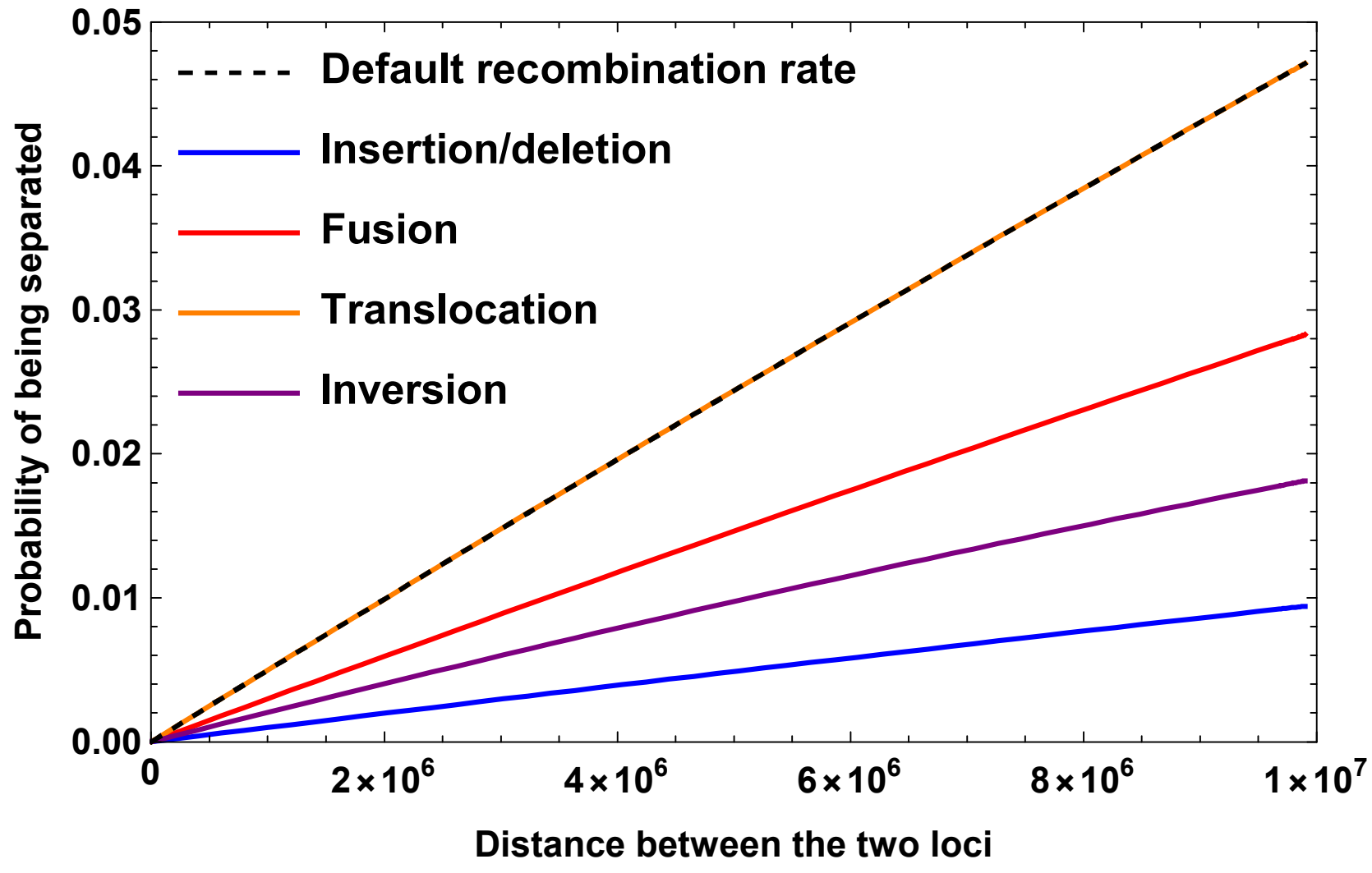

Illustration of the main expressions when the loci are close to each other

```
With[{βDSB = 10^-8, ϕCO = 0.5, ϕGC = 0.7, λ = 50, fAa = 0.4, fAA = 0.2,
  h = 0.8, βNDJ = .5, y0 = 1, y1 = 10000000, x1 = 100000, c = 0.8}, LogLogPlot[
  {Prec[βDSB, ϕCO, ϕGC, λ, x1, x1 + d], PIndel[βDSB, ϕCO, ϕGC, λ, fAA, x1, x1 + d],
   PFusion[βDSB, ϕCO, ϕGC, λ, fAA, fAa, h, βNDJ, x1, x1 + d],
   PTranslocation[βDSB, ϕCO, ϕGC, λ, fAa, c, βNDJ, x1, x1 + d], PInversion[βDSB,
    ϕCO, ϕGC, λ, fAa, y0, y1, x1, x1 + d], Prec[βDSB, ϕCO, ϕGC, λ, x1, x1 + d]},
  {d, 10, y1 - x1}, PlotRange → {{10, 10^4}, {10^-8, 0.0001}},
  PlotStyle → {{Dashed, Black}, Blue, Red, Orange, Purple, {Dashed, Black}},
  PlotLegends → Placed[{"Default recombination rate", "Insertion/deletion",
     "Fusion", "Translocation", "Inversion"}, {Left, Top}],
  BaseStyle → {Thick, Bold, 12}, ImageSize → 450, Frame → True,
  FrameLabel → {"Distance between the two loci", "Probability of being separated"}]]
```

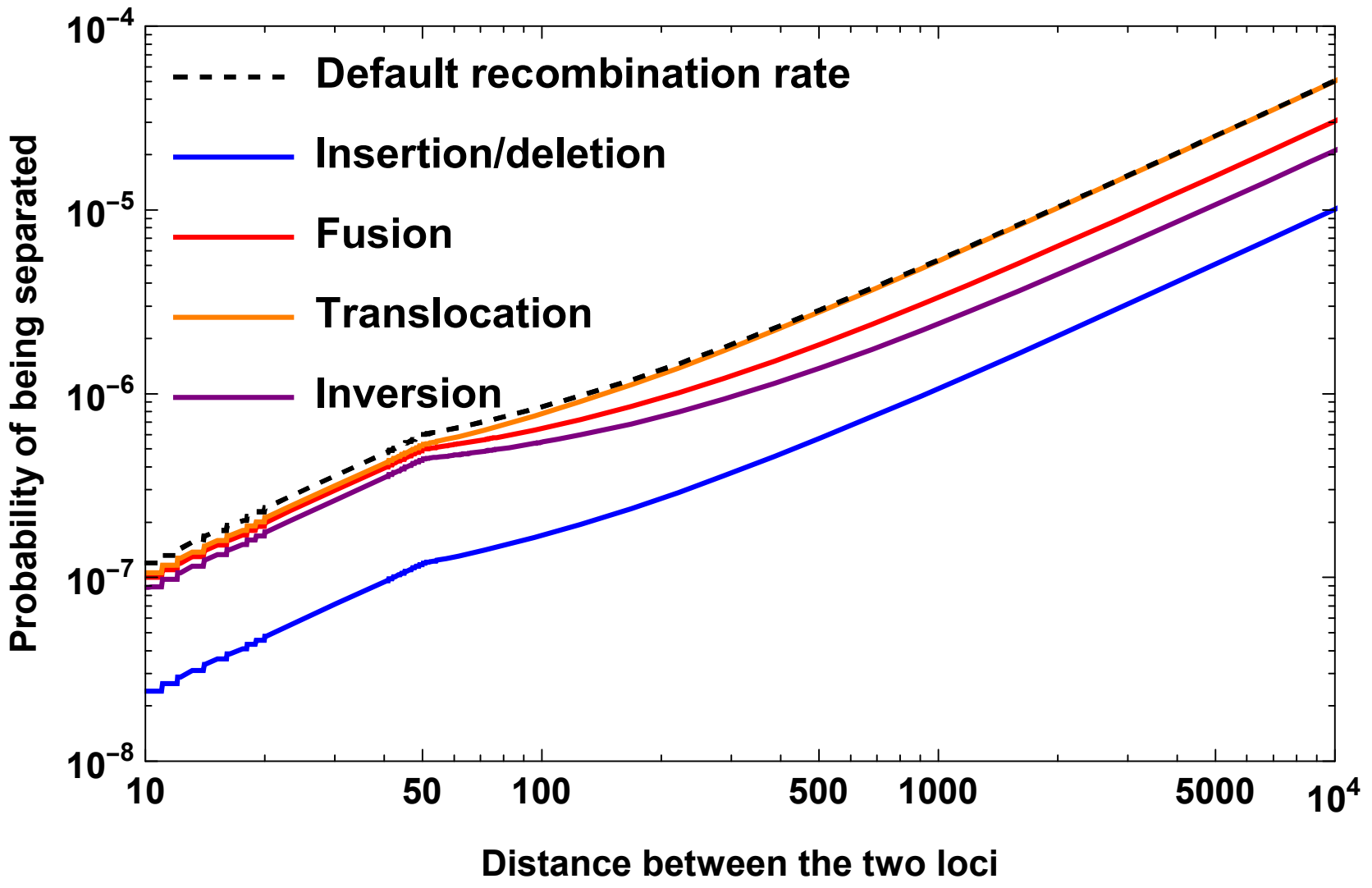

Approximations given in the main text

```
With[{βDSB = 10^-8, ϕCO = 0.5, ϕGC = 0.7, λ = 50, fAa = 0.4,
  fAA = 0.2, h = 0.8, βNDJ = .5, y0 = 1, y1 = 10000000, x1 = 100000, c = 0.8},
 Plot[{(-x1 ϕCO + (x1 + d) ϕCO + λ ϕGC) βDSB, (-x1 ϕCO + (x1 + d) ϕCO + λ ϕGC) βDSB fAA,
   (((-d) (1 - (fAa + fAA) h + fAa (-1 + h) βNDJ) ϕCO) / (-1 + fAa βNDJ) + λ ϕGC) βDSB,
   (-x1 ϕCO + (x1 + d) ϕCO + ((-1 + fAa (c + βNDJ - c βNDJ)) λ ϕGC)/(-1 + fAa βNDJ)) βDSB,
   (-((-1 + 2 fAa) (-d) ϕCO)/(-1 + fAa) + λ ϕGC) βDSB, (-x1 ϕCO + (x1 + d) ϕCO + λ ϕGC) βDSB},
  {d, 50, y1 - x1}, PlotRange → {{50, 10^7}, {10^-8, 0.05}},
  PlotStyle → {{Dashed, Black}, Blue, Red, Orange, Purple, {Dashed, Black}},
  PlotLegends → Placed[{"Default recombination rate", "Insertion/deletion",
     "Fusion", "Translocation", "Inversion"}, {Left, Top}],
  BaseStyle → {Thick, Bold, 12}, ImageSize → 450, Frame → True,
  FrameLabel → {"Distance between the two loci", "Probability of being separated"}]]
```

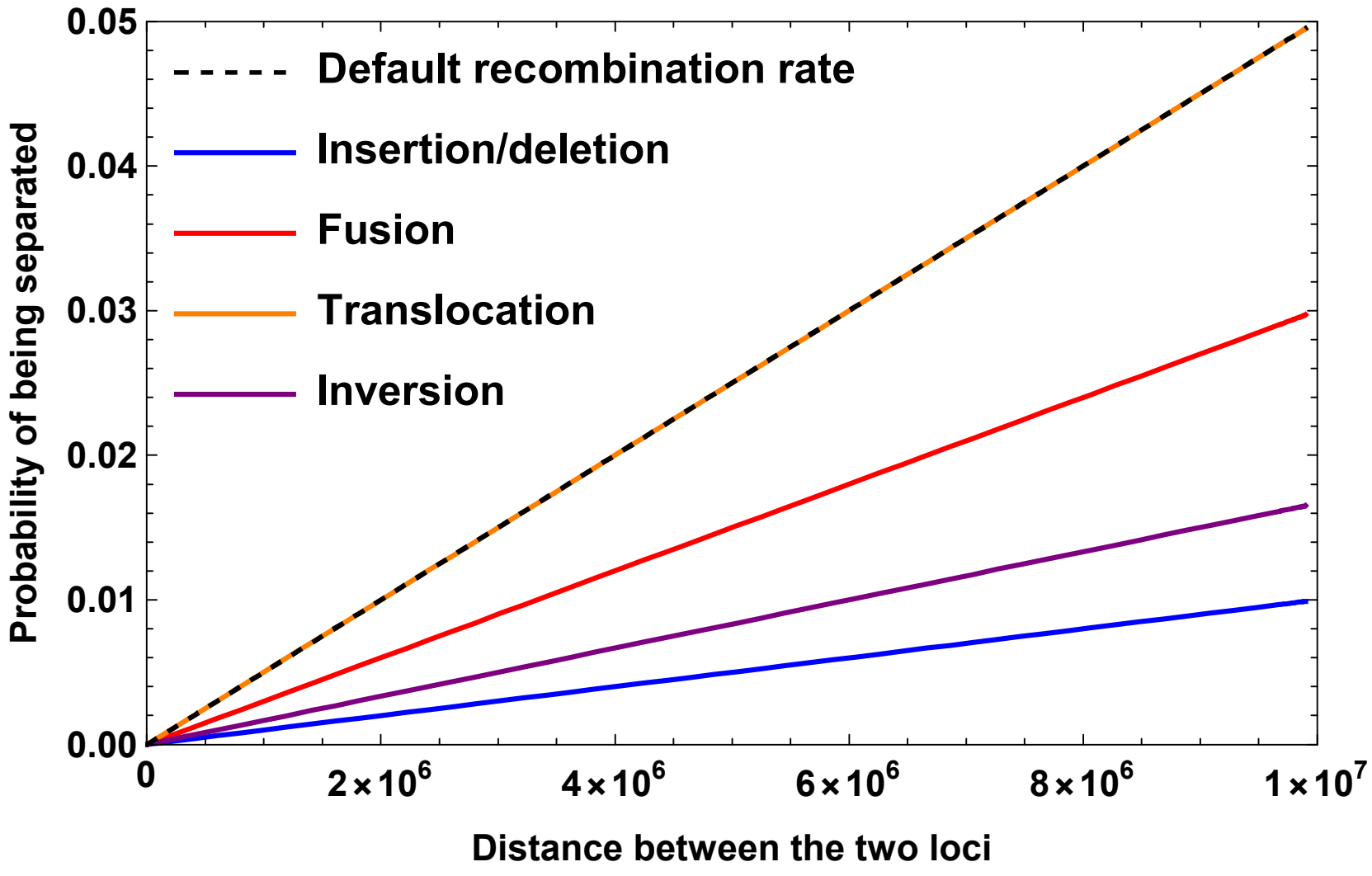

Approximations if the two loci are close to each other

```
With[{βDSB = 10^-8, ϕCO = 0.5, ϕGC = 0.7, λ = 50, fAa = 0.4,
  fAA = 0.2, h = 0.8, βNDJ = .5, y0 = 1, y1 = 10 000 000, x1 = 100 000, c = 0.8},
 LogLogPlot[{(-x1 ϕCO + (x1 + d) ϕCO + λ ϕGC) βDSB, (-x1 ϕCO + (x1 + d) ϕCO + λ ϕGC) βDSB fAA,
   (((-d) (1 - (fAa + fAA) h + fAa (-1 + h) βNDJ) ϕCO) / (-1 + fAa βNDJ) + λ ϕGC) βDSB,
   (-x1 ϕCO + (x1 + d) ϕCO + ((-1 + fAa (c + βNDJ - c βNDJ)) λ ϕGC) / (-1 + fAa βNDJ)) βDSB,
   (-((-1 + 2 fAa) (-d) ϕCO) / (-1 + fAa) + λ ϕGC) βDSB, (-x1 ϕCO + (x1 + d) ϕCO + λ ϕGC) βDSB},
  {d, 50, y1 - x1}, PlotRange → {{50, 10^4}, {10^-7, 0.0001}},
  PlotStyle → {{Dashed, Black}, Blue, Red, Orange, Purple, {Dashed, Black}},
  PlotLegends → Placed[{"Default recombination rate", "Insertion/deletion",
     "Fusion", "Translocation", "Inversion"}, {Left, Top}],
  BaseStyle → {Thick, Bold, 12}, ImageSize → 450, Frame → True,
  FrameLabel → {"Distance between the two loci", "Probability of being separated"}]]
```

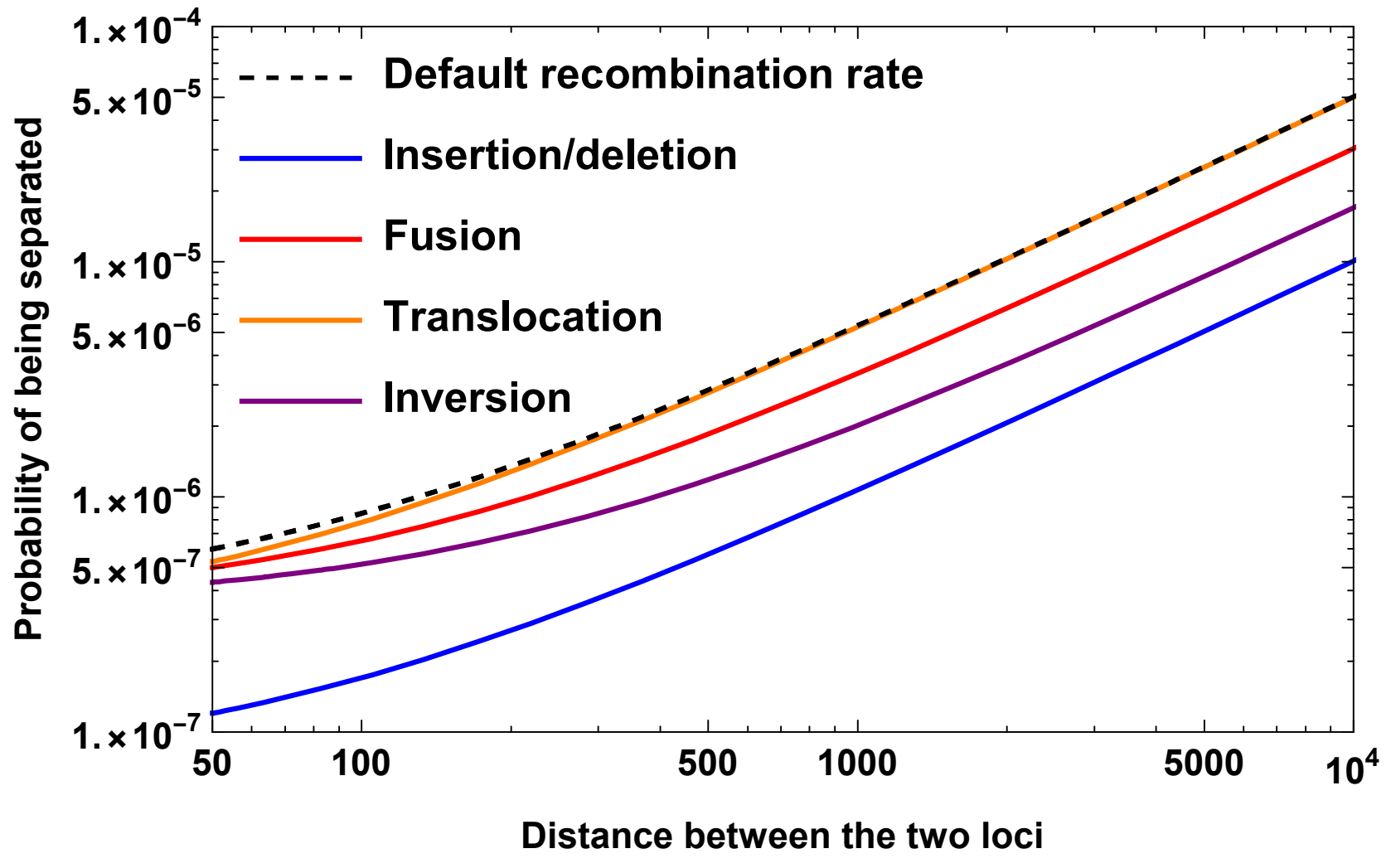

Evaluation of the factor S1 and S2 given in the main document:

```
With[{βDSB = 10^-8, ϕCO = 0.5, ϕGC = 0.7, λ = 50, fAa = 0.4,
  fAA = 0.2, h = 0.8, βNDJ = .5, y0 = 1, y1 = 10 000 000, x1 = 100 000, c = 0.8},
 {((1 - (fAa + fAA) h + fAa (-1 + h) βNDJ) / (-1 + fAa βNDJ)), (-((-1 + fAa (c + βNDJ - c βNDJ)) / (-1 + fAa βNDJ)))}]
```

```
{-0.6, -0.8}
```

Illustration with the expressions and their approximations

```
With[{βDSB = 10^-8, ϕCO = 0.5, ϕGC = 0.7, λ = 50, fAa = 0.4,
  fAA = 0.2, h = 0.8, βNDJ = .5, y0 = 1, y1 = 10000000, x1 = 100000, c = 0.8},
 Plot[{Prec[βDSB, ϕCO, ϕGC, λ, x1, x1 + d], PIndel[βDSB, ϕCO, ϕGC, λ, fAA, x1, x1 + d],
   PFusion[βDSB, ϕCO, ϕGC, λ, fAA, fAa, h, βNDJ, x1, x1 + d],
   PTranslocation[βDSB, ϕCO, ϕGC, λ, fAa, c, βNDJ, x1, x1 + d], PInversion[βDSB,
    ϕCO, ϕGC, λ, fAa, y0, y1, x1, x1 + d], Prec[βDSB, ϕCO, ϕGC, λ, x1, x1 + d],
   (-x1 ϕCO + (x1 + d) ϕCO + λ ϕGC) βDSB, (-x1 ϕCO + (x1 + d) ϕCO + λ ϕGC) βDSB fAA,
   (((-d) (1 - (fAa + fAA) h + fAa (-1 + h) βNDJ) ϕCO) / (-1 + fAa βNDJ) + λ ϕGC) βDSB,
   (-x1 ϕCO + (x1 + d) ϕCO + ((-1 + fAa (c + βNDJ - c βNDJ)) λ ϕGC) / (-1 + fAa βNDJ)) βDSB,
   (-((-1 + 2 fAa) (-d) ϕCO) / (-1 + fAa) + λ ϕGC) βDSB, (-x1 ϕCO + (x1 + d) ϕCO + λ ϕGC) βDSB},
  {d, 10, y1 - x1}, PlotRange → {{50, 9.9 × 10^6}, {10^-8, 0.05}}, PlotStyle →
   {{Dashed, Black}, Blue, Red, Orange, Purple, {Dashed, Black}, {Dotted, Black},
    {Dotted, Blue}, {Dotted, Red}, {Dotted, Orange}, {Dotted, Purple}, {Dotted, Black}},
  PlotLegends → Placed[{"Default recombination rate", "Insertion/deletion",
     "Fusion", "Translocation", "Inversion"}, {Left, Top}],
  BaseStyle → {Thick, Bold, 12}, ImageSize → 450, Frame → True,
  FrameLabel → {"Distance between the two loci", "Probability of being separated"}]]
```

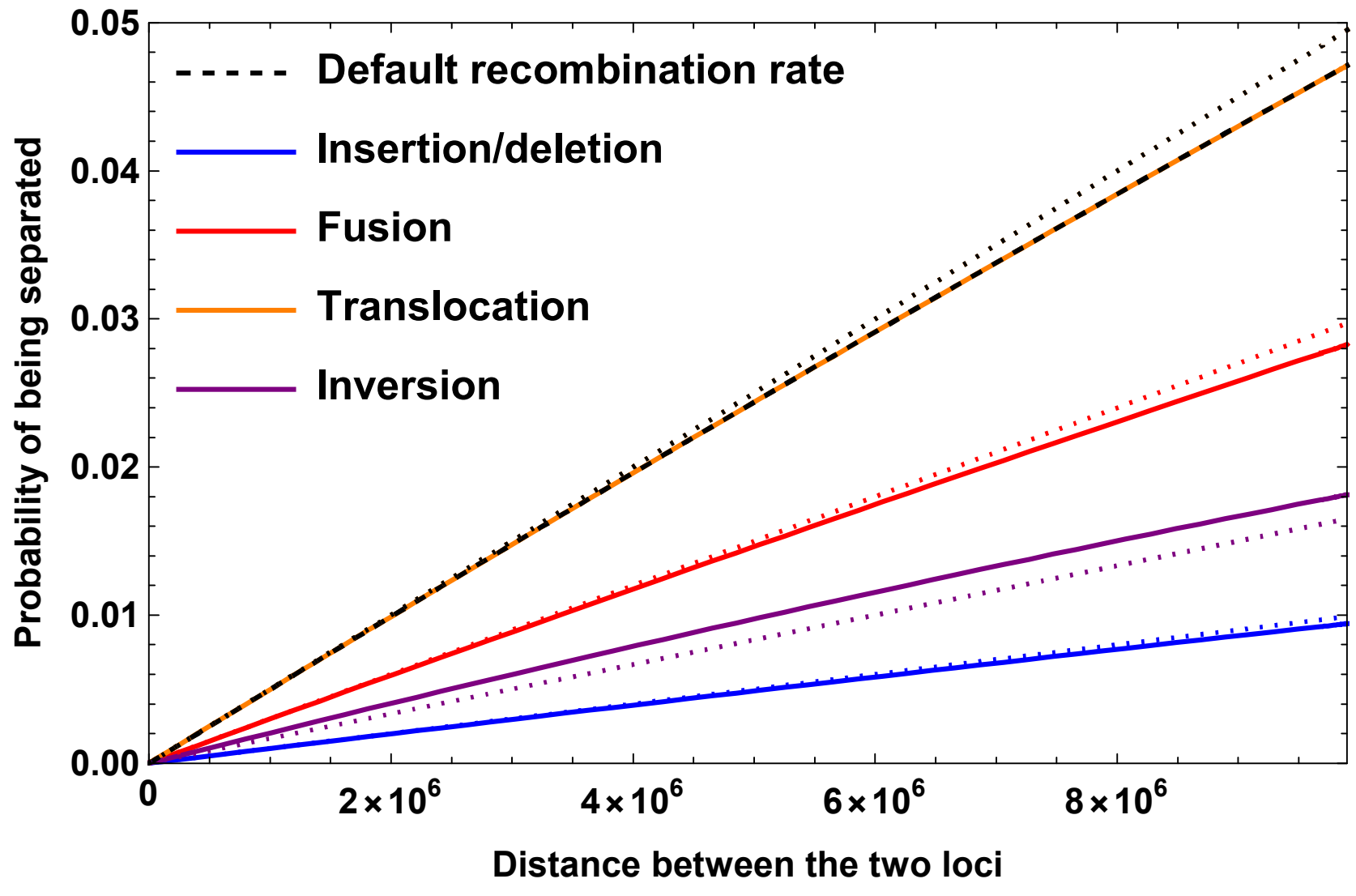

Illustration with the expressions and their approximations if the two loci are close to each other :

```
With[{βDSB = 10^-8, ϕCO = 0.5, ϕGC = 0.7, λ = 50, fAa = 0.4, fAA = 0.2,
  h = 0.8, βNDJ = .5, y0 = 1, y1 = 10000000, x1 = 100000, c = 0.8}, LogLogPlot[
  {Prec[βDSB, ϕCO, ϕGC, λ, x1, x1 + d], PIndel[βDSB, ϕCO, ϕGC, λ, fAA, x1, x1 + d],
   PFusion[βDSB, ϕCO, ϕGC, λ, fAA, fAa, h, βNDJ, x1, x1 + d],
   PTranslocation[βDSB, ϕCO, ϕGC, λ, fAa, c, βNDJ, x1, x1 + d],
   PInversion[βDSB, ϕCO, ϕGC, λ, fAa, y0, y1, x1, x1 + d],
   Prec[βDSB, ϕCO, ϕGC, λ, x1, x1 + d], If[d ≤ λ, (d ϕCO + d ϕGC) βDSB, (d ϕCO + λ ϕGC) βDSB],
   If[d ≤ λ, (d fAA ϕCO + d fAA ϕGC) βDSB, (d fAA ϕCO + fAA λ ϕGC) βDSB], If[d ≤ λ,
    (d (-ϕCO + fAa h ϕCO + fAA h ϕCO + fAa βNDJ ϕCO - fAa h βNDJ ϕCO - ϕGC + fAa βNDJ ϕGC) βDSB) /
     (-1 + fAa βNDJ), ((-d ϕCO + d fAa h ϕCO + d fAA h ϕCO + d fAa βNDJ ϕCO -
         d fAa h βNDJ ϕCO - λ ϕGC + fAa βNDJ λ ϕGC) βDSB) / (-1 + fAa βNDJ)], If[d ≤ λ,
    ((-d ϕCO + d fAa βNDJ ϕCO - d ϕGC + c d fAa ϕGC + d fAa βNDJ ϕGC - c d fAa βNDJ ϕGC) βDSB) /
     (-1 + fAa βNDJ),
    ((-d ϕCO + d fAa βNDJ ϕCO - λ ϕGC + c fAa λ ϕGC + fAa βNDJ λ ϕGC - c fAa βNDJ λ ϕGC) βDSB) /
     (-1 + fAa βNDJ)], If[d ≤ λ, (d (ϕCO - 2 fAa ϕCO + ϕGC - fAa ϕGC) βDSB) / (1 - fAa),
    ((-d ϕCO + 2 d fAa ϕCO - λ ϕGC + fAa λ ϕGC) βDSB) / (-1 + fAa)],
   If[d ≤ λ, (d ϕCO + d ϕGC) βDSB, (d ϕCO + λ ϕGC) βDSB],
   If[d ≤ λ, (d ϕCO + d ϕGC) βDSB, (d ϕCO + λ ϕGC) βDSB]},
  {d, 10, y1 - x1}, PlotRange → {{10, 10^4}, {10^-8, 0.0001}},
  PlotStyle → {{Dashed, Black}, Blue, Red, Orange, Purple,
    {Dashed, Black}, {Dotted, Black}, {Dotted, Blue}, {Dotted, Red},
    {Dotted, Orange}, {Dotted, Purple}, {Dotted, Black}},
  PlotLegends → Placed[{"Default recombination rate", "Insertion/deletion",
     "Fusion", "Translocation", "Inversion"}, {Left, Top}],
  BaseStyle → {Thick, Bold, 12}, ImageSize → 450, Frame → True,
  FrameLabel → {"Distance between the two loci", "Probability of being separated"}]]
```

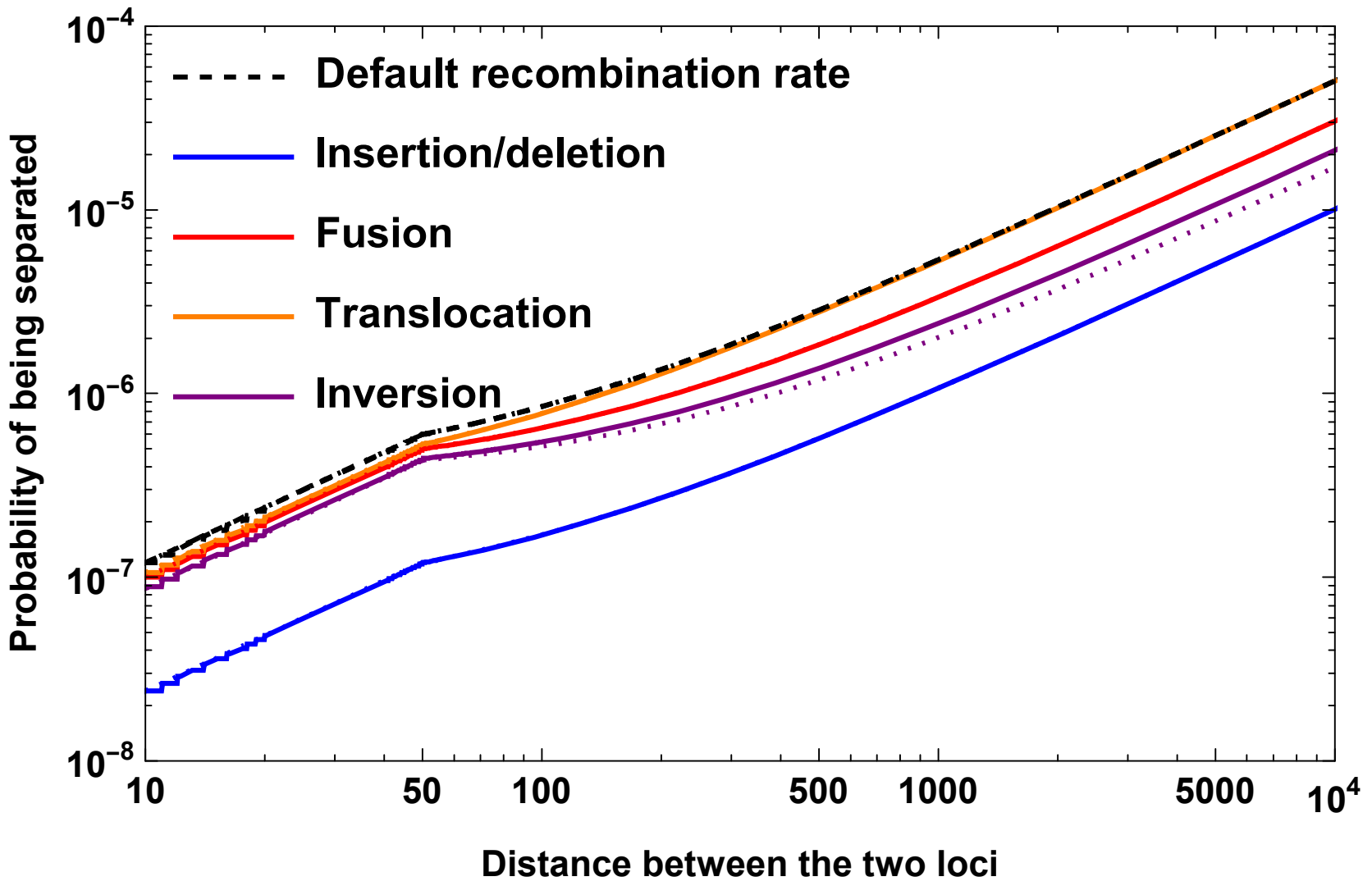